\documentclass[a4paper]{aa}

\usepackage{txfonts}

\usepackage{graphicx, epsfig}
\usepackage{multirow}
\usepackage{natbib}
\bibpunct{(}{)}{;}{a}{}{,} 

\usepackage{url}
\begin{document}

\title{Multi-frequency monitoring of $\gamma$-ray loud blazars}

\subtitle{I. Light curves and spectral energy distributions}

\author{U.~Bach\inst{1}\thanks{Current address: Max-Planck-Institut f\"ur
Radioastronomie, Auf dem
H\"ugel 69, 53121 Bonn, Germany}
\and C.~M.~Raiteri\inst{1}
\and M.~Villata\inst{1}
\and L.~Fuhrmann\inst{1,2\,\star}
\and C.~S.~Buemi\inst{3}
\and V.~M~Larionov\inst{4}
\and P.~Leto\inst{5}
\and A.~A.~Arkharov\inst{6}
\and J.~M.~Coloma\inst{7}
\and A.~Di~Paola\inst{8}
\and M.~Dolci\inst{9}
\and N.~Efimova\inst{6,4}
\and E.~Forn\'e\inst{7}
\and M.~A.~Ibrahimov\inst{10}
\and V.~Hagen-Thorn\inst{4}
\and T.~Konstantinova\inst{4}
\and E.~Kopatskaya\inst{4}
\and L.~Lanteri\inst{1}
\and O.~M.~Kurtanidze\inst{11}
\and G.~Maccaferri\inst{12}
\and M.~G.~Nikolashvili\inst{11}
\and A.~Orlati\inst{12}
\and J.~A.~Ros\inst{7}
\and G.~Tosti\inst{2}
\and C.~Trigilio\inst{3}
\and G.~Umana\inst{3}
} 

\offprints{U.~Bach,~\email{ubach@mpifr-bonn.mpg.de}} 

\institute{INAF, Osservatorio Astronomico di Torino, Via Osservatorio 20, 
10025 Pino Torinese (TO), Italy 
\and Osservatorio Astronomico, Universit\`a di Perugia, via B. Bonfigli, 06126
Perugia, Italy
\and INAF, Osservatorio Astrofisico di Catania, Via S. Sofia 78, 95123
Catania, Italy
\and Astronomical Institute, St. Petersburg State University, Universitetsky pr.
28, Petrodvoretz, 198504 St. Petersburg, Russia
\and INAF, Istituto di Radioastronomia, Sezione di Noto, C.da Renna Bassa - Loc.
Case di Mezzo C.P. 141, 96017 Noto, Italy 
\and Pulkovo Astronomical Observatory of the Russian Academy
of Sciences, Russia
\and Agrupaci\'o Astron\`omica de Sabadell, PO Box 50, 08200 Sabadell, Spain
\and INAF, Osservatorio Astronomico di Roma, Via di Frascati 33, 00040 Monte
Porzio Catone, Italy 
\and INAF - Osservatorio Astronomico di Collurania Teramo, Via Maggini snc, 64100
Teramo, Italy
\and Ulugh Beg Astronomical Institute, 33 Astronomical Str., Tashkent 700052,
Uzbekistan
\and Abastumani Astrophysical Observatory, 383762 Abastumani, Georgia
\and INAF, Istituto di Radioastronomia, Sezione di Medicina, Via Fiorentina
3508/B, 40059 Medicina, Italy
} 

\date{Received 13 October 2006; Accepted 16 November 2006}

\titlerunning{Multi-frequency monitoring of $\gamma$-ray loud blazars}

\authorrunning{U.\ Bach et al.}

\abstract{Being dominated by non-thermal emission from aligned relativistic
jets, blazars allow us to elucidate the physics of extragalactic jets, and,
ultimately, how the energy is extracted from the central black hole in
radio-loud active galactic nuclei.}
{Crucial information is provided by broad-band spectral energy distributions (SEDs),
their trends with luminosity and correlated
multi-frequency variability. With this study we plan to obtain a database of
contemporaneous radio-to-optical spectra of a sample of blazars,
which are and will be observed by current and future high-energy satellites.}
{Since December 2004 we are performing a monthly multi-frequency radio monitoring of a
sample of 35 blazars at the antennas in Medicina and Noto. Contemporaneous
near-IR and optical observations for all our observing epochs are organised.}
{Until June 2006 about 4000 radio measurements and 5500 near-IR and optical measurements
were obtained. Most of the sources show significant variability in all observing
bands. Here we present the multi-frequency data acquired during the first eighteen months
of the project, and construct the SEDs for the best-sampled sources.} {}

\keywords{galaxies: active -- galaxies: BL Lacertae objects: 
general -- galaxies: jets -- galaxies: quasars: general}

\maketitle

\section{Introduction}\label{sec:intro}

Blazars form a sub-group of radio-loud active galactic nuclei (AGN) showing
extreme variability at all wavelengths, high degrees of linear polarization, strong
$\gamma$-ray emission, apparent superluminal motion of radio jet components, and
brightness temperatures exceeding the Compton limit (see e.g.\
\citealt{1999APh....11..159U}).  They include BL\,Lac objects as well as quasars
with flat radio spectrum and violent  variability in the optical band. The large
amount of work done in the last decades led to the commonly accepted scenario in
which a rotating supermassive black hole  surrounded by an accretion disk with an
intense plasma jet closely aligned to the line of sight is responsible for the
blazar emission. Relativistic electrons produce the soft photons through
synchrotron emission, while hard photons are likely produced by inverse-Compton
scattering. This overall scenario, however, still presents a large number of poorly
understood details which, in turn, lead to a wide variety of models and call for
long-term and multi-wavelength campaigns capable of providing the necessary
observational constraints.

From the time scales of variability  key information on the AGN structure can be
derived, down to linear scales or flux density levels not accessible even with
interferometric imaging, but the mechanisms which are  responsible for variability
are not well understood yet. Different models are discussed in the literature such
as shock-in-jets (e.g.\
\citealt{1985ApJ...298..114M,1985ApJ...298..296A,1996etrg.conf...45M})  or
colliding relativistic plasma shells (e.g.\
\citealt{2001MNRAS.325.1559S,2004A&A...421..877G}). Further, in the case of
precessing binary black-hole systems (e.g.\
\citealt{1980Natur.287..307B,1988ApJ...325..628S}), rotating helical jets (e.g.\
\citealt{1998MNRAS.293L..13V,1999A&A...347...30V}) or helical trajectories of
plasma elements (e.g.\ \citealt{1992A&A...255...59C}), the models suggest that
changes in the direction of forward beaming introduces flares due to the lighthouse
effect. Thus, variability furnishes important clues into size, structure, physics
and dynamics of the radiating source region. A tool to study the radiating
particles is provided by the analysis of the synchrotron continuum over a range of
frequencies as broad as possible. The shape of the synchrotron spectrum gives
direct insight into the shape of the electron energy distribution, thus also
constraining the emission by the inverse-Compton process at other wavelengths.

The most extensive blazar monitoring campaigns have been carried out at radio
frequencies. The University of Michigan monitoring program at 4.8, 8.0 and
14.5\,GHz has obtained data on over 200 sources for over three decades
(\citealt{2003AAS...202.1801A}) and is still continued. At higher frequencies, the
Mets\"ahovi Radio Observatory has reported observations at 22, 37 and 87\,GHz of
157 extragalactic radio sources
(\citealt{1998A&AS..132..305T,2004A&A...427..769T,2005A&A...440..409T}), many of
which have been monitored for over 20 years, but the monitoring is now continued
only at 37\,GHz. The Bologna group (\citealt{1996A&AS..120...89B}) has observed at
408\,MHz 125 radio sources from 1975 to 1990. A sample of 550 sources is monitored
at six radio frequencies from 1.4 to 31\,GHz quasi simultaneously since many years
at the RATAN-600 telescope (\citealt{2002PASA...19...83K}).

However, aside from some organised broad multi-wavelength campaigns, most of our 
previous studies of the spectral energy distributions (SEDs) of blazars were
performed by combining measurements carried out non-simultaneously (over a period
of one or more months), using the above mentioned radio surveys and data from
various optical, near-IR, and X-ray observatories (e.g.\
\citealt{1999A&A...352...19R,2001A&A...377..396R,2003A&A...402..151R,2004A&A...424..497V}).
Since blazars are highly variable, it is therewith necessary to extrapolate between
non simultaneous points, which is not an ideal procedure. Hence, it is important to
take simultaneous broad-band spectra that are not affected by the
variability. Therefore, we started a monthly multi-frequency radio monitoring of a
sample of 35 $\gamma$-ray loud blazars at the antennas in Medicina and Noto, and
organised contemporaneous near-IR and optical observations for most of the
observing epochs. This provides a valuable database of time-dependent SEDs,
which can be used to test different jet emission models. The monitoring is supposed
to support also the next Italian high-energy space missions AGILE, which will
observe many of our monitored sources.

In this article we will present the light curves and SEDs of those sources which
have been monitored for at least one year by the end of June 2006. Further
publications with a more detailed timing analysis and modelling of the SEDs are in
preparation. Throughout this paper we will assume a flat universe model with the
following parameters: Hubble constant $H_0=71$\,km\,s$^{-1}$\,Mpc$^{-1}$, a
pressure-less matter content $\Omega_{\rm m}=0.3$, and a cosmological constant
$\Omega_{\rm \lambda}=0.7$ (\citealt{2003ApJS..148..175S}). Cosmology-dependent
values quoted from other authors are scaled to these parameters.

The paper is organised as follows: Section~\ref{sec:sample} introduces the source
sample. In Sect.~\ref{sec:observations} we will describe the data acquisition and
reduction procedures and in Sect.~\ref{sec:results} we will present the resulting
light curves and SEDs. Peculiar variability behaviours or special treatments during
the data analysis for single sources are described in Sect~\ref{sec:individ}. 
Conclusions are drawn in Sect.~\ref{sec:summary}.

\section{The source sample}\label{sec:sample}

At the beginning of our monitoring we started with a sample of $\gamma$-ray loud
sources which had already been monitored in the optical band at the Perugia and
Torino observatories since 1994 and which were relatively bright also in the radio
band (S$_{\rm 1.4GHz}>1$Jy). In view of the forthcoming launch of the high-energy
satellites AGILE and GLAST, we subsequently added some more sources from the third
EGRET catalog (\citealt{1999ApJS..123...79H}) with declination $>-15^\circ$, which
are also bright at optical ($B<19$) and radio wavelengths (S$_{\rm 1.4\,GHz}>1$\,Jy).
The 33 sources of the present sample are listed in Table~\ref{tab:sample}. In this
paper we present the data of those sources which were monitored for at least one
year (marked in bold face in the table). The remaining ones will be reported in a
later paper (Bach et al.\ in prep.).

\begin{table}[htbp]
\caption{The complete source list. Given are IAU name, position, type
(Quasar, BL\,Lac, or Seyfert) and redshift . The sources presented in this paper are
marked in bold face, the others will be presented in a later paper.} 
\hspace{-3mm}
  \begin{tabular}{llccc}
\hline
IAU & Other& R.A.~\&~Dec.\ (J2000) & Type & $z$\\
Name & Name & [h:m:s]~~~[d:m:s] & &\\
\hline
{\bf 0219+428}   & 3C\,66A       &  02:22:39.6   +43:02:08  & Q & 0.44 \\
{\bf 0235+164}   &AO\,0235+16    &  02:38:38.9   +16:36:59  & Q & 0.94 \\
0336$-$019       &CTA\,026       &  03:39:30.9 $-$01:46:36  & Q & 0.85 \\
{\bf 0420$-$014} &PKS 0420$-$01  &  04:23:15.8 $-$01:20:33  & Q & 0.91 \\
{\bf 0440$-$003} &NRAO\,190      &  04:42:38.6 $-$00:17:43  & Q & 0.84 \\
{\bf 0528+134}   &PKS\,0528+134  &  05:30:56.4   +13:31:55  & Q & 2.06 \\
{\bf 0716+714}   &S5\,0716+71    &  07:21:53.4   +71:20:36  &BL & $>$0.3 \\
{\bf 0735+178}   &PKS\,0735+17   &  07:38:07.3   +17:42:19  &BL & 0.42 \\
0736+017         &PKS\,0736+01   &  07:39:18.0   +01:37:05  & Q & 0.19 \\
0827+243         & OJ\,248       &  08:30:52.1   +24:11:00  & Q & 0.94 \\
0829+046         & OJ\,49        &  08:31:48.9   +04:29:38  &BL & 0.17 \\
{\bf 0836+710}   &S5\,0836+71    &  08:41:24.3   +70:53:42  & Q & 2.17 \\
{\bf 0851+202}   & OJ\,287       &  08:54:48.8   +20:06:31  &BL & 0.31 \\
{\bf 0954+658}   &S4\,0954+65    &  09:58:47.2   +65:33:55  &BL & 0.37 \\
{\bf 1156+295}   & 4C\,+29.45    &  11:59:31.8   +29:14:44  & Q & 0.73 \\
1219+285         & W\,Com        &  12:21:31.7   +28:13:58  &BL & 0.10 \\
{\bf 1226+023}   & 3C\,273       &  12:29:06.7   +02:03:09  & Q & 0.16 \\
{\bf 1253$-$055} & 3C\,279       &  12:56:11.1 $-$05:47:21  &BL & 0.54 \\
1334$-$127       &PKS\,1335$-$127&  13:37:39.8 $-$12:57:25  &BL & 0.54 \\
1354+195         &4C\,+19.44     &  13:57:04.4   +19:19:07  &Sy & 0.72 \\
{\bf 1510$-$089} &PKS 1510$-$08  &  15:12:50.5 $-$09:06:00  & Q & 0.36 \\
1606+105         &4C\,+10.45     &  16:08:46.1   +10:29:08  & Q & 1.23 \\
{\bf 1611+343}   & DA\,406       &  16:13:41.0   +34:12:48  & Q & 1.40 \\
{\bf 1633+382}   &S4\,1633+38    &  16:35:15.4   +38:08:04  & Q & 1.81 \\
{\bf 1641+399}   & 3C\,345       &  16:42:58.8   +39:48:37  & Q & 0.59 \\
{\bf 1730$-$130} &NRAO\,530      &  17:33:02.6 $-$13:04:49  & Q & 0.90 \\
{\bf 1741$-$038} &PKS\,1741$-$03 &  17:43:58.8 $-$03:50:05  & Q & 1.05 \\
1807+698         & 3C\,371       &  18:06:50.7   +69:49:28  &BL & 0.05 \\
{\bf 2200+420}   & BL\,Lac       &  22:02:43.2   +42:16:40  &BL & 0.07 \\
2223$-$052       & 3C\,446       &  22:25:47.2 $-$04:57:01  &BL & 1.40 \\
{\bf 2230+114}   &CTA\,102       &  22:32:36.4   +11:43:50  & Q & 1.04 \\
{\bf 2251+158}   & 3C\,454.3     &  22:53:57.7   +16:08:54  & Q & 0.86 \\
2344+092         & 4C\,09.74     &  23:46:36.8   +09:30:45  & Q & 0.68 \\
\hline
  \end{tabular}
 \label{tab:sample}
\end{table}

\section{Observations and data reduction}\label{sec:observations}

\subsection{Radio data}

Radio measurements were taken at the 32\,m radio antennas of the INAF-Istituto di
Radioastronomia in Medicina and Noto on a monthly basis at frequencies of 5, 8, and
22\,GHz. At the beginning all frequencies (5, 8, and 22\,GHz) were taken at both
telescopes, but because of the better performance of the Noto antenna at high
frequencies, the observations were split to 5 and 8\,GHz at Medicina and 22\,GHz at
Noto after the first 6 months. Unfortunately, the antenna in Noto was out of
operation for the second half of 2005, because of a technical failure. During this
time observations at 22\,GHz were occasionally performed at Medicina. Because
blazars tend to be more variable at higher radio frequencies and a new 43\,GHz
receiver at Noto became available, from early 2006 data are acquired at 43\,GHz in
Noto (Bach et al.\ in prep.), while the radio monitoring at 22\,GHz is done with
the Medicina antenna only. 

The nominal system equivalent flux densities (SEFD) of the Medicina antenna are
296\,Jy, 284\,Jy, and 1318\,Jy at 5, 8, and 22\,GHz respectively, and 260\,Jy,
770\,Jy, and 800\,Jy at the Noto antenna. On both telescopes the data is taken from
the VLBI backend. Because of radio interference (RFI) the bandwidth at 5\,GHz is
limited to 32\,MHz at Medicina, but at the other frequencies the full IF
(intermediate frequency channel) of 400\,MHz bandwidth is used.

The observations in Medicina are performed using ON/OFF measurements through the
VLBI field system. The ON/OFFs are obtained by pointing the telescope subsequently
ON the source and five beam widths to the east and west OFF the source. A typical
ON/OFF sequence (8 to 20 ON/OFF subscans) takes about 8 to 18 minutes, depending on
the frequency and source flux, and corresponds to 40 to 100 seconds ON-source
integration time. Regular sky dip measurements (every 3 to 4 hours) are used to
correct the data for opacity effects. The 22 and 43\,GHz flux measurements in Noto
are obtained with a similar method but the ON/OFF subscans are realized by
``chopping'' the secondary reflector between the ON and OFF position, which allows
a faster switching. Here, a typical ON/OFF sequence of 20 subscans with 1 second
integration time takes about 3 minutes. At the lower frequencies (5 and 8\,GHz) the
antenna performs raster scans in right ascension and flux densities are obtained by
Gaussian fitting of the right ascension profiles. Opacity measurements are taken
every half  an hour.

Since the VLBI field system is used to perform the observations in Medicina and
Noto, SNAP-command scripts or schedules are used to control the telescope. Most of
the observations at Medicina are carried out remotely from Torino. Usually a single
schedule is used for each 2--3 day observing run, switching between frequencies
every few hours, but in case of technical problems or dramatic weather changes it
is possible to change the schedule at any time. All the information provided by the
observing system (e.g.: weather information, system temperature, as well as the
counts from the ON/OFF measurements) are written in ASCI format into a log-file.
The field system also performs an analysis of the ON/OFF scans and provides
uncalibrated antenna temperatures. However, to be more flexible in the data editing
and to discard also single subscans, we wrote a Python program which extracts the
information from the log-file and analyses again the individual subscans. The
resulting table contains Stoke I antenna temperatures averaged over the two IFs
containing the left and right circular polarized signals. The software which was
used during a previous blazar radio monitoring project performed in Medicina and
Noto between 1996 and 2000 (\citealt{2001A&A...379..755V}) was not usable anymore,
because the ON/OFF procedure of the field system has changed in the meantime. Noto
measurements from both observing methods are written in separate log-files and are
reduced using IDL programs.

The flux density measurements are calibrated using a set of standard primary
calibrators, namely 3C\,123, 3C\,147, 3C\,286, 3C\,295, 3C\,48, and NGC\,7027, and
in addition at 22\,GHz DR\,21 and W3OH. Their flux densities were calculated by the
polynomial expressions given by \cite{1977A&A....61...99B} and
\cite{1994A&A...284..331O}. Presumably the most stable calibrator is 3C\,286,  for
which 7.32\,Jy (5\,GHz), 5.22\,Jy (8\,GHz), and 2.50\,Jy (22\,GHz) were adopted for
the calibration, but at 22\,GHz mainly NGC\,7027 (5.58\,Jy) and DR\,21 (19.04\,Jy)
were used, because they are more reliable (less variable) than quasars at this
frequency. The calibrators are also used to obtain elevation-dependent gain curves
to correct the data for this effect. The accuracy of the ON/OFF measurements in
Medicina at 5 and 8\,GHz is usually about 5\% to 10\%, depending on the source flux
density and weather conditions. At 22\,GHz the accuracy is more weather-dependent,
and ranges between 10\% to 15\% during good weather, but can be much worse during
unstable weather conditions. Raster scans, like they are used in Noto, yield a
slightly better accuracy, the typical error at 5 and 8\,GHz being  2--5\%. At
22\,GHz the typical error at the Noto antenna is about 10\%.

\subsection{Optical and near-IR data}

Contemporaneously to the radio observations we organise optical ($BVRI$ bands) and
near-IR ($JHK$ bands) observations at several  observatories, which are listed in
Table~\ref{tab:obslog}. The optical and near-IR magnitudes are obtained by 
differential photometry to known reference stars\footnote{The adopted photometry
can be found at \url{http://www.to.astro.it/blazars/webt/bamp/list.html}.}.

The images were reduced using standard procedures for photometry. To avoid offsets
due to different calibration choices, the measurements from the different
observatories were collected as instrumental magnitudes of the source and
references stars and were calibrated all together. The uncertainty was calculated
from the scattering of the references stars, which are supposed to be not variable.

\begin{table}[htbp]
\caption{The participating observatories. Given are the name, size of the
telescope, observing bands, and number of observations. Note that the data of
Crimea and St.\ Petersburg are counted together, since they were observed by the
same working group and were reduced all together using the same procedures. The
uppercase numbers at the observatories denote the corresponding  institutes (see
page~\pageref{sec:intro}).}
\centering
  \begin{tabular}{lrcc}
\hline
Observatory & \multicolumn{1}{c}{Size [m]} & \multicolumn{1}{c}{Bands} & \multicolumn{1}{c}{N}\\
\hline
Medicina$^{12}$ & 32.0 & 5, 8, 22\,GHz & 2953\\
Noto$^5$ & 32.0 & 5, 8, 22\,GHz & 1405\\
Maidanak (T-60)$^{10}$ & 0.6 & $R$ & 3327\\
Abastumani$^{11}$ & 0.7 & $BR$ & 342\\
\vspace{-3pt}
Crimea$^4$ & 0.7 & $BVRI$ & \multirow{2}{11pt}{791}\\
St.\ Petersburg$^4$ & 0.4 & $BVRI$ & \\
Perugia$^2$ & 0.4 & $VRI$ & 80\\
Campo Imperatore (AZT-24)$^9$ & 1.1 & $JHK$ & 2898\\
Torino (OTAP)$^1$ & 1.0 & $BVRI$ & 1726\\
Sabadell$^7$ & 0.5 & $R$ & 204\\
\hline
  \end{tabular}
 \label{tab:obslog}
\end{table}

Flux densities were derived from magnitudes using the absolute calibrations of
\cite{1998A&A...333..231B}, after dereddening according to the extinction laws of
\cite{1989ApJ...345..245C}. We adopted the $B$-band values of Galactic extinction
provided by Nasa/IPAC extragalactic
database\footnote{\url{http://nedwww.ipac.caltech.edu}} (NED), from
\cite{1998ApJ...500..525S}. Non standard procedures that were used to reduce the
data of 0235+164 and 2200+420 (BL\,Lac) are described in Sect.~\ref{sec:individ}.

\section{Results}\label{sec:results}

In this section we present the radio as well as the near-IR and optical light
curves  of the best sampled sources. The corresponding SEDs constructed with
simultaneous data are also shown. Peculiarities in the light curves of single
objects are described in the next section (Sect.~\ref{sec:individ})

\subsection{Radio light curves}

Radio light curves at 5, 8, and 22\,GHz of those blazars that are monitored from
the beginning of our project are shown in Fig.~\ref{fig:lc}. As visible in the
legend, different symbols refer to different frequencies, whereas open symbols
denote data taken in Medicina and filled symbols denote data from the antenna in
Noto. The very good agreement between the radio measurements from the antennas in
Noto and Medicina, taken with different methods and reduced by different authors,
confirms a overall good quality of the data. During some epochs the data was
affected by bad weather conditions, which results in larger scattering and  error
bars, specially at 22\,GHz.

Most of the sources display significant variability on monthly time scales. Some
sources show an achromatic behaviour, where the flux densities at different
frequencies vary simultaneously, e.g.\ 1510-089. In other cases variations at
22\,GHz are likely leading those at the lower frequencies, e.g.\ 1226+023 (3C\,273)
and 2251+158 (3C\,454.3). In the case of 2200+420 (BL\,Lacertae), we recognise both
characteristics. A flare that is more pronounced at 22\,GHz appeared in early 2005
and a second one equally bright at all frequencies followed about 250 days later.

\subsection{Optical light curves}\label{sec:olc}

Optical light curves are not shown for all sources, but only for those which have a
good time coverage (Fig.~\ref{fig:olc}). The lack of data is mostly due to solar
conjunction or bad weather periods. The measurements from different observatories
agree quite well and the variations in the optical $BVRI$ bands and near-IR  $JHK$
bands usually show similar amplitudes. To allow the reader to better
distinguish between different bands only $B$, $R$, $J$, and $K$ light curves
are shown. Measurements in $V$, $I$, and $H$ are shown in the SEDs only (next
Sect.). In general, the sources are all more variable in the optical bands than at
radio wavelengths. In Fig.~\ref{fig:olc} we complement the light curves of 0235+164
and 2251+158 with data taken by the Whole Earth Blazar Telescope\footnote{\tt
http://www.to.astro.it/blazars/webt/}  (WEBT) during specific campaigns (grey
symbols, see \citealt{Raiteri2006} for 0235+164 and \citealt{2006A&A...453..817V}
for 2251+258). 

Part of the data presented here were also provided to other multi-wavelength
campaigns and the specific time intervals are marked by yellow (grey in the printed
version) vertical bars in the figure. For 0851+202 the periods of two short-term
WEBT campaigns organized in conjunction with two pointings of the X-ray satellite
{XMM-Newton} (Ciprini et al.\ in prep.) are highlighted. Similarly, the yellow
strip crossing the 1253-055 light curves indicates  the time of the January--March
2006 WEBT campaign on this object (B\"ottcher et al.\ in  prep.). In the case of
2251+158, the yellow area corresponds to the new WEBT campaigns following the
2004--2005 one (Villata et al. in prep; Raiteri et al. in prep.).

\subsection{Spectral energy distributions}\label{sec:sed}

In Fig.~\ref{fig:SEDs} the SEDs of several blazars are shown.  Grey dots represent
non-simultaneous literature data taken from NED, while different symbols (and
colours) have been used to distinguish the contemporaneous data acquired for this
project.  The sources shown here are those for which several epochs of simultaneous
data exist and whose SEDs show significant variability. In this energy range, we
expect to see the synchrotron component peaking in the sub-millimetre to infrared
region, in the case of these low-energy peaked blazars.

The lines connecting the data points are third-order polynomial fits, and do not
represent any emission model.  Although the SEDs of blazars seem to be represented
quite well by log-parabolic fits, as demonstrated by various authors
(\citealt{Landau1986,Perri2003,Massaro2004,Massaro2004a}), we used third-order fits
because they show more clearly the optical spectral changes which are present in
some sources. In those cases where a ``blue bump'' appears, we used cubic spline
interpolations. Thus the lines are drawn just to guide the eye and to better
illustrate the behaviour of the source. A proper modelling of the SED is behind the
scope of this article, but will be presented in a future publication. Therefore,
the peaks of the SEDs in Fig.~\ref{fig:SEDs} might not exactly match the position
where most of the energy is released, but should provide an estimate of the true
location.

As already seen in the radio light curves, also the broad-band variability can be
either chromatic or achromatic. The long-term variability of the BL\,Lac object
0716+714 seems to affect the emission from the radio to the optical wavelengths at
the same time,  whereas the behaviour of 2251+158 is completely different. Indeed,
in 2005 the source experienced a historical flare, peaking around May in the
optical band \citep{2006A&A...445L...1F,2006A&A...449L..21P,2006A&A...453..817V},
and possibly a bit later in the near-IR one, while it appeared at 22\,GHz with a
few months of delay, in August--September 2005 (see Fig.~\ref{fig:lc}).

Although most blazars of our sample reach their emission peak in the 
millimetre-to-infrared regime and their SED declines in the optical band, some
sources show different behaviours. Exceptions are e.g.\  0716+714, whose SED
appears rather flat in the optical band, and the quasars
0836+710, 1226+026, and 1510-089, whose optical part of the SED is increasing
again (see the individual paragraphs in Sect.~\ref{sec:individ} for more details).
Commonly, this increase is referred to as a ``blue bump", which could be due to
emission from an accretion disk (e.g.\ \citealt{1990A&ARv...2..125B}). 

For some sources, e.g.\ 0235+164, 1156+195, and 2251+158, we can also see spectral
changes during different states in the optical, whose dependence on the brightness
level, if any, is not clear. In the other cases the colours do not change much
during different brightness levels.

\section{Notes on individual sources}\label{sec:individ}

\paragraph{\bf 0219+428 (3C\,66A)} This BL\,Lac object has been observed intensively
at optical wavelengths (e.g.\ \citealt{2005ApJ...631..169B}) and
\cite{1999ApJ...521..561L} found a possible periodicity of 65 days in its optical
bright state. Very fast superluminal motion of up to 29\,$c$ was reported by
\cite{2001ApJS..134..181J}. Despite some small flux changes, the radio light curves
reveal not much variability (Fig.~\ref{fig:lc}). A major problem for radio
measurements of 3C\,66A is its close companion 3C\,66B, which at some position
angles appears in the OFF measurements or the cross scans and makes the calibration
more difficult. During the first  $\sim 50$~days of the optical light curve a
brightness decrease is visible, which reached its minimum at about JD=2453650
(October 2005), with some shorter variability on top of the long-term trend
(Fig.~\ref{fig:olc}).

\paragraph{\bf 0235+164} This BL\,Lac object usually displays strong variability at
optical and radio wavelength. Although very compact, sometimes jet features with
high superluminal motion can be observed. \cite{2001ApJS..134..181J} reported an
apparent speed as fast as 40\,$c$, and another feature at about 26\, $c$ was
observed by \cite{2006ApJ...640..196P}. An outburst in early 2004 was expected as
part  of a $\sim 5.7$ year quasi-periodicity resulting from the analysis  of the
radio (and optical) historical light curves (\citealt{2001A&A...377..396R}).  A
long-term multi-wavelength campaign was organized by the WEBT to monitor the
predicted outburst (\citealt{2005A&A...438...39R,2006A&A...452..845R,Raiteri2006}).
The grey symbols in Fig.~\ref{fig:olc} represent the data taken in the last period
of the WEBT campaign. Only two arcseconds south of the source, there is another
AGN, which significantly affects the optical flux measurements of the 0235+164.
Therefore, the magnitudes shown in Fig.~\ref{fig:olc} have been corrected for this
contribution. Moreover, the near-IR and optical fluxes shown in Fig.~\ref{fig:SEDs}
were obtained  by taking into account not only the Galactic extinction, but also
the extinction due to the foreground galaxy (see details in
\citealt{2005A&A...438...39R}).  We observed the source in a very low state. The
light curves show a slow increase at radio frequencies, reaching a maximum at
22\,GHz around JD=2453700 (November 2005) and several weeks later at 8 and 5\,GHz.
Some significant variability is also visible at the optical wavelengths.

\paragraph{\bf 0716+714} This BL\,Lac object is one of the best-studied intraday
variable sources. Simultaneous radio, optical, UV, and X-ray monitoring yielded a
short duty cycle of variability at all frequencies, as well as a correlation
between the rapid variations at different frequency regimes
(\citealt{1996AJ....111.2187W}). Those measurements suggested very small sources
sizes and high brightness temperatures exceeding $10^{17}$\,K.  The data collected
during  a recent large multi-wavelength campaign allowed to obtain more moderate
values  of $10^{14}$ to $10^{16}$\,K (\citealt{2006A&A...451..797O,Agudo2006};
Fuhrmann et al.\ in prep.), well in agreement with the Doppler factors derived from
kinematical studies of the jet (e.g,
\citealt{2001ApJS..134..181J,2004ApJ...609..539K,2005A&A...433..815B}) and space
VLBI measurements of the source size (\citealt{2006A&A...452...83B}). A detailed
study of the multi-wavelength long-term variability was performed by
\cite{2003A&A...402..151R}. During the present monitoring, 0716+714 showed three
flares in the radio, lasting about 100 days, and exhibiting a smaller amplitude
than those occurred in the past. But in our observing period, the radio flux also
went down to its historical minimum of about 0.5\,Jy (Fig.~\ref{fig:lc}). Similar
variations appear also in the optical band (Fig.~\ref{fig:olc}), and an inspection
of the SEDs confirms that the optical and radio flux densities vary
contemporaneously (Fig.~\ref{fig:SEDs}).  The fact that the peak of the SEDs falls
at higher energies compared to many of the other blazars in our sample confirms the
classification of this source as an  ``intermediate" BL\,Lac (as 3C\,66A), half-way
between the low-energy peaked (LBL) and the high-energy peaked (HBL) BL\,Lacs.

\paragraph{\bf 0836+710} This ultra-luminous quasar at $z=2.17$ is showing a well
correlated broad-band variability that also seems to correlate with the structural
changes seen in the VLBI jet (e.g.\ \citealt{1998A&A...334..489O}). The source
showed only weak variability (Figs.~\ref{fig:lc}~\&~\ref{fig:olc}). The hard
near-IR--optical spectrum implies that the blue bump component, likely due to
thermal emission from the accretion disc, is still dominant in the near-IR band,
and consequently it is not possible to distinguish the high-energy part of the
synchrotron component. This is actually expected, because of the high redshift of
the source.

\paragraph{\bf 0851+202 (OJ\,287)} This well-studied BL\,Lac object is known for
its $\sim 12$-year periodicity in the optical light curve
(\citealt{1988ApJ...325..628S,1996A&A...315L..13S}). The outbursts appear as
double-peaked and their periodic occurrence is mostly interpreted in  terms of a
binary black hole system
(\citealt{1988ApJ...325..628S,1996ApJ...460..207L,1998MNRAS.293L..13V,2000ApJ...531..744V}).
Depending on the model, different dates around 2006 are predicted for the next
outburst, but the recent strong activity makes an  identification not easy.
According to \cite{2006ApJ...646...36V,2006ApJ...643L...9V}, the outburst occurred
in late 2005 may have already been the predicted one, but a final statement is not
possible before 2008, when all predicted dates will have passed. During our radio
monitoring we observed an increasing trend, with some flares detectable at 22\,GHz
(Fig.~\ref{fig:lc}).  The optical light curves show the declining phase of the
large outburst around November 2005, and another flare around April--May 2006
(Fig.~\ref{fig:olc}). The SEDs shows that the amplitude of the optical variations
of OJ\,287 were slightly larger than in the radio (Fig.~\ref{fig:SEDs}).

\paragraph{\bf 0954+658} This BL\,Lac object is another famous IDV source
(\citealt{1993A&A...271..344W,1995ARA&A..33..163W,2000MNRAS.315..229G}). The jet
structure shows many bends from milliarcsecond to arcsecond scales and speeds of up
to 9\,$c$ were measured with VLBI
(\citealt{1996MNRAS.283..759G,1992AJ....104.1687K}). The long-term radio light
curve (Fig.~\ref{fig:lc}) displays some variability, which is more pronounced at
22\,GHz, while greater activity characterizes the optical behaviour
(Fig.~\ref{fig:olc}).  This becomes even more evident in the SEDs, where in
addition one can notice the flattening of the optical spectrum during the flare of
January 2005 (Fig.~\ref{fig:SEDs}).

\paragraph{\bf 1156+295} This quasar is extremely variable over all the
electromagnetic spectrum, from radio waves to $\gamma$-rays
\citep{1983ApJ...274...62W,1992ApJ...398..454W,1999ApJS..123...79H}. 1156+295 has a
straight arcsecond scale jet and a more curved jet on milliarcsecond scales, with
jet components moving at superluminal speeds of up to 16\,$c$ (e.g.\
\citealt{2004A&A...417..887H}). Our observations reveal a major optical outburst in
March 2006 (Fig.~\ref{fig:olc}). An increasing trend is also visible in the radio
light (Fig.~\ref{fig:lc}) and it will be interesting to see if we can find a
connection between the two events.

\paragraph{\bf 1226+023 (3C\,273)}  This object shows all the characteristics that
are typical of high-luminosity quasars: a flat radio spectrum of the core, strong
and rapid variability in all the observed energy ranges (e.g.\
\citealt{1997ApJ...483..161V,1999MNRAS.310..571M}), variable polarization, and a
radio jet with superluminal motion (e.g.\ \citealt{1981Natur.290..365P}).
Additionally, it shows an optical and X-ray jet, and a very prominent UV excess,
the so-called big blue bump (e.g.\
\citealt{1996A&A...314..414R,2005A&A...431..477J}). In the SEDs the contribution of
this component begins to dominate over the synchrotron one in the optical band
(Fig.~\ref{fig:SEDs}). Since it is likely due to thermal radiation from the
accretion disc, it is not expected to be strongly variable on short time
scales, and indeed the optical light curves do not show important flux changes over the
whole time range (Fig.~\ref{fig:olc}). On the contrary, a noticeable flux increase
has recently been observed at 22\,GHz, which might propagate also to the lower
radio frequencies in the next future (Fig.~\ref{fig:lc}).

\paragraph{\bf 1253$-$055 (3C\,279)} This quasar, considered the first superluminal
source (\citealt{1971Sci...173..225W}), is often nearly as bright as 3C\,273 at
high frequencies. Since it is also a bright $\gamma$-ray source
(\citealt{1999ApJS..123...79H}), 3C\,279 has been the target of a number of
campaigns of contemporaneous monitoring from $\gamma$-ray to radio frequencies
(e.g.\ \citealt{1996ApJ...459...73G,1998ApJ...497..178W}).  Another campaign has
recently been performed, which was motivated by an optical bright state observed in
December 2005--January 2006, which in turn triggered ToO observations by {\it
Chandra} and {\it INTEGRAL} (B\"ottcher et al.\ in prep.). The radio light curve at
22\,GHz shows a noticeable decrease in the first months of the project, and a more
moderate increase afterwards; on the contrary, the radio flux at 8 and 5\,GHz  did
not change very much (Fig.~\ref{fig:lc}). Due to the low declination of the source,
the optical light curve has a large gap around the solar conjunction
(Fig.~\ref{fig:olc}). The SEDs show a fast changing optical spectrum
(Fig.~\ref{fig:SEDs}).

\paragraph{\bf 1510$-$089} This is another extreme quasar, highly polarized,  and
bright at X-ray and $\gamma$-ray energies.  It exhibits a highly bent radio jet
structure with fast superluminal speeds of up to 20\,$c$
(\citealt{1997ApJ...491..515S,1999ApJS..123...79H,2001ApJ...549..840H}). In the
radio band, we saw the source fading simultaneously at all frequencies
(Fig.~\ref{fig:lc}). Noticeable variations were seen also in the near-IR
(Fig.~\ref{fig:olc}). Note also that our SEDs (Fig.~\ref{fig:SEDs}) show a dip
between the near-IR and optical bands, in agreement with the observations by
\cite{1979ApJ...230...79N} in 1977 (data reported by NED, grey dots). In that case,
however, the optical spectrum was quite harder and it seems that a much stronger
emission component was present towards smaller wavelength. Such an increase in the
spectrum of a quasar is commonly referred to thermal emission from an accretion
disk (e.g.\ \citealt{1990A&ARv...2..125B}), but in that case, although the
observations are separated by $\sim30$~years, the component should not be so strongly
variable. A similar behaviour was observed in AO\,0235+164, where a variable
UV-to-soft-X-ray bump was found
(\citealt{2005A&A...438...39R,2006A&A...452..845R,Raiteri2006}). In this case the
authors speculate that the variability could be due to a strong change of e.g.\ the
accretion rate. Another possibility would be the presence of another emission
component, e.g.\ synchrotron radiation from an inner jet region with respect to
that producing the radio--near-IR radiation. For 1510$-$089 the latter explanation
would be favoured by the noticeable optical variability also observed on short time
scales (e.g.\ \citealt{1998A&AS..127..445R}). Further multi-wavelength  monitoring
will help to better characterize the behaviour of this component.

\paragraph{\bf 1641+399 (3C\,345)} The quasar 3C\,345 has been studied extensively
at all accessible wavelength. The observed strong variability in the optical and
radio bands, the existence of possible periodicities, and the detection of bent
trajectories of the radio jet components, made this source particularly interesting
to test jet precession and binary black hole models (e.g.\
\citealt{1986ApJ...308...93B,Caproni2004,Lobanov2005,1999ApJ...521..509L,Schramm1993,Stevens1996,Webb1994}.
Our radio light curves are slowly declining (Fig.~\ref{fig:lc}), but in the optical
band a strong increase can be seen from the beginning of the last observing season
(Fig.~\ref{fig:olc}).  This becomes even more evident by looking at the evolution
of the SEDs (Fig.~\ref{fig:SEDs}), specially when considering that the last four
SEDs are separated by only a few weeks.

\paragraph{\bf 2200+420 (BL\,Lacertae)}  This is the prototype of the BL\,Lac
objects, which are well-known for their pronounced variability at all wavelengths,
from the radio to the $\gamma$-ray band. BL\,Lac has been studied intensively since
its discovery and was the target of several multi-wavelengths campaigns
(\citealt{1997ApJ...490L.145B,1999ApJ...515..140S,1999ApJ...521..145M,2002A&A...390..407V,2003ApJ...596..847B,2004A&A...424..497V,2004A&A...421..103V}).
The radio emission appears well correlated, where variations at higher frequencies
lead the lower-frequency ones by several days to a few months (e.g.\
\citealt{2004A&A...424..497V}). The detection of a fair correlation between the
optical variations and the radio variability from the VLBI-core suggests a common
origin in the inner portion of the jet (\citealt{Bach2006}). A number of
superluminal jet components have been observed displaying bent trajectories and
speeds from 3\,$c$ to $9\,c$ (e.g.\
\citealt{1990ApJ...352...81M,2000ApJS..129...61D,2003MNRAS.341..405S,2004ApJ...609..539K,2005AJ....130.1418J}).
During our monitoring BL\,Lac showed its typical variability behaviour
(Figs.~\ref{fig:lc}~\&~\ref{fig:olc}). There are flares which are more pronounced
at higher frequencies (including optical counter parts) and appear delayed at lower
frequencies, and flares that are occurring contemporaneously and with comparable
amplitudes, but appear only at radio wavelength. To construct the SED the optical
flux densities of BL\,Lac were corrected for the contribution of the host galaxy
using the $R$-band value by \cite{2000ApJ...532..740S} ($R=15.55\pm0.02$) and
deriving the other bands from the colour indices for elliptical galaxies with
$M_{\rm V}<-21$ from \cite{2001MNRAS.326..745M}.

\paragraph{\bf 2230+114 (CTA\,102)}  This quasar shows the typical blazar behaviour
and appears in the source lists of many radio and optical monitoring
projects. Multi-wavelength monitoring and imaging revealed correlated variability
and radio jet speeds of up to 21\,$c$ were derived from VLBI observations (e.g.\
\citealt{2003A&A...405..473R}). Our near-IR--optical light curve is dominated by a
huge flare lasting more than a hundred days around October 2005
(Fig.~\ref{fig:olc}), and we can follow now an ongoing radio flare
(Fig.~\ref{fig:lc}). The SEDs show impressively the change of the spectrum with the
propagation of the flare towards lower frequencies (Fig.~\ref{fig:SEDs}), similarly
to the case of 2251+158 (see next paragraph).

\paragraph{\bf 2251+158 (3C\,454.3)} This is another well-known quasar and also one
of the brightest ones. The source is showing remarkable variability at all
wavelengths, high degrees of polarization, superluminal motion of the radio jet
components, and a X-ray jet
(\citealt{1984ApJ...286..503C,1987Natur.328..778P,1993ApJ...407L..41H}). With the
start of our monitoring 3C\,454.3 underwent an exceptional optical outburst lasting
for more than one year, subsequently followed by a radio flare
(Figs.~\ref{fig:lc}~\&~\ref{fig:olc} and
\citealt{2006A&A...445L...1F,2006A&A...453..817V}). The maximum brightness detected
was $R=12.0$, which, at $z=0.859$, represents the most luminous quasar state
observed until today ($M_B \sim -31.4$, \citealt{2006A&A...453..817V}).  Therefore,
the SED shows a remarkable variability and the propagation of the flare from higher
to lower energies is clearly visible (Fig.~\ref{fig:SEDs}). At the beginning of the
rising phase the near-IR--optical spectrum appears rather flat and changes to a
very steep one after the brightness peak was reached. The radio flare did not reach
frequencies much below 15\,GHz  and, although some structural changes in the
radio jet were detected, so far no jet component related to the outburst is
visible (Marscher et al.\ 2007, in prep). A detailed analysis of the
radio--optical correlations  will be presented in an upcoming paper (Villata et
al.\ in prep.). 

\section{Conclusions}\label{sec:summary}

During the first 18 months of our multi-wavelength monitoring we obtained about 4000
data points in the radio bands, at 5, 8, and 22\,GHz, and 5500 near-IR and optical
measurements. 

Some of our sources were targets of observations by X-ray satellites and by
ground-based TeV telescopes, and we contributed data to the multi-wavelength WEBT
campaigns on 0235+164 \citep{Raiteri2006}, OJ\,287 (Ciprini et al.\ in prep.),
3C\,279 (B\"ottcher et al.\ in prep.), BL\,Lac  (Villata et al.\ in prep.), and
3C\,454.3  (\citealt{2006A&A...453..817V};  Villata et al.\ in prep; Raiteri et
al.\ in prep.).

Most of the sources display significant variability on monthly time scales. Some
sources show an achromatic variability behaviour, where the frequencies vary almost
simultaneously, as e.g.\, 1510-089.  In other cases the flux changes at the higher
frequencies are likely leading lower-frequency ones, as for 1226+023 (3C\,273) and
2251+158 (3C\,454.3),  where flares are visible at 22\,GHz, but are not yet visible
at 8 and 5\,GHz. In the case of 2200+420 (BL\,Lac) we find  both characteristics. A
flare that is more pronounced at 22\,GHz appeared in early 2005, and a second one
equally bright at all frequencies appeared about 250 days later. 

Many possible mechanisms which are able to explain the various variability
behaviours are discussed in the literature (see Sect.~\ref{sec:intro} for
references). They manly divide into geometrical models where changes of the viewing
angle lead to variations of the Doppler beaming factor and models which propose
changes of the physical parameters of the jet plasma itself. Because of the limited
time interval covered by the present project, it is not possible yet to favour or
rule out some of these models. Neither does it allow to perform a meaningful
statistical analysis on the time scales of variability or on cross correlations
among flux changes at different frequencies. However, the planned continuation of
this multi-wavelength monitoring during the next years is expected to provide a wide
database for our targets that will allow us to extract more information on the
mechanisms at the origin of blazar emission variability.

Moreover, with the launch of AGILE and GLAST the GeV energy domain will be
accessible and our monitoring will support the high-energy observations with 
low-energy data, to learn more about the inter-connection between the synchrotron
and inverse-Compton components of blazar SEDs.

\begin{acknowledgements}
We thank the referee, Esko Valtaoja, for his suggestions on improving the paper. We
thank the staff at the Medicina and Noto radio observatories for their help and
support during this monitoring. This work is based on observations with the
Medicina and Noto telescopes operated by INAF - Istituto di Radioastronomia. AZT-24
telescope is operated under agreement between Pulkovo, Rome and Teramo
observatories. This work is supported by the European  Community's Human Potential
Programme under contract HPRCN-CT-2002-00321 (ENIGMA Network). St.Petersburg
University  team acknowledges support from Russian Federal Program for Basic
Research under grant 05-02-17562. This research has made use of the NASA/IPAC
Extragalactic Database (NED) which is operated by the Jet Propulsion Laboratory,
California Institute of Technology, under contract with the National Aeronautics
and Space Administration.

\end{acknowledgements}

\bibliographystyle{aa} 
\bibliography{references}

\begin{figure*}[htbp]
\centering
\includegraphics[bb= 48 49 619 801,angle=0,width=16cm,clip]{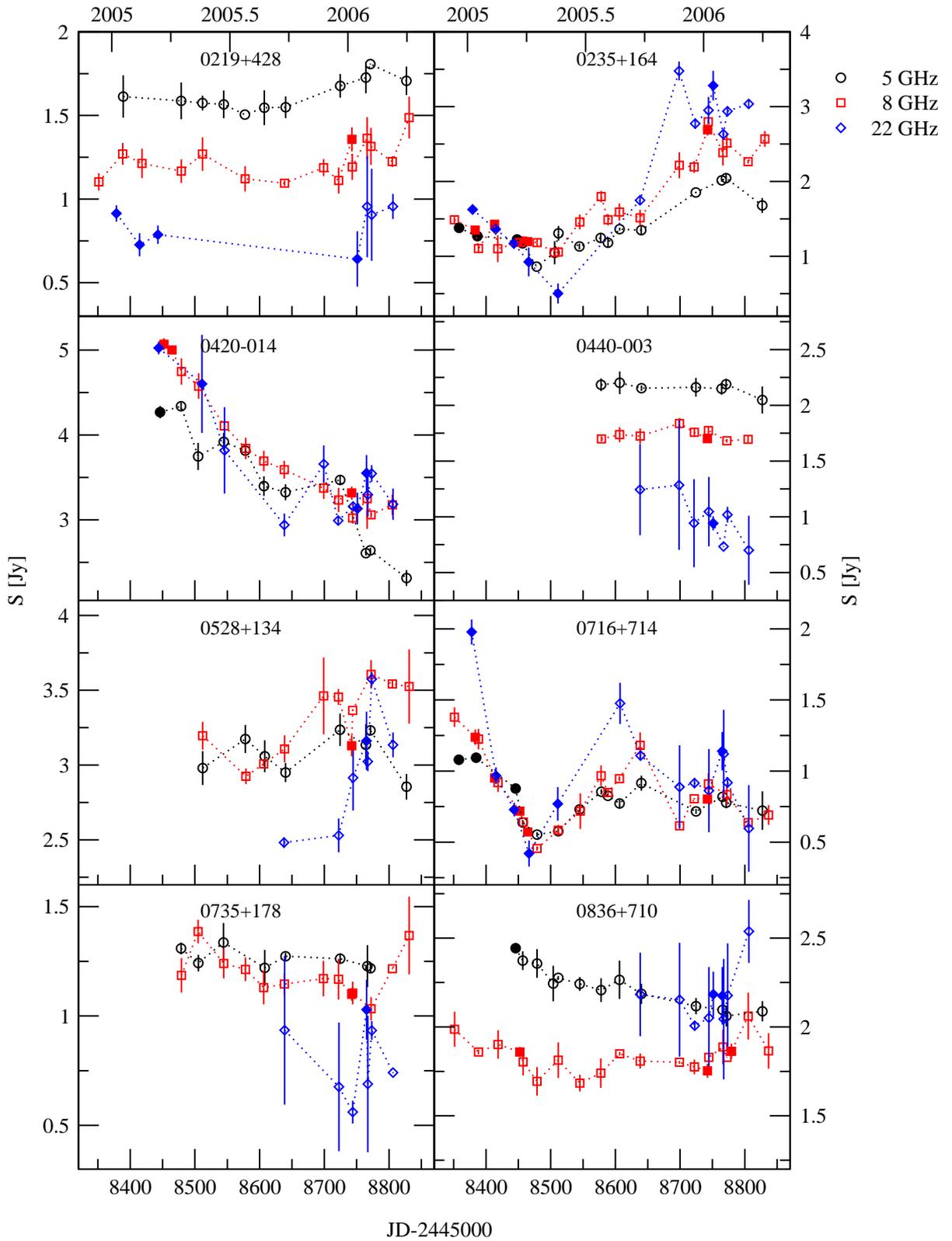}
\caption{Radio light curves at 5, 8, and 22\,GHz of our blazar sample between
December 2004 and June 2006 taken with the antennas in Medicina (open
symbols) and Noto (filled symbols). Although some of the 22\,GHz measurements
suffer from bad weather conditions, we are able to trace the variability
behaviour very nicely. In most cases the flares appear first at the higher
frequency, suggesting that the variability is caused by a shock propagating
along the jet (continued on next page).}\label{fig:lc}
\end{figure*}

\setcounter{figure}{0}
\begin{figure*}[htbp]
\centering
\includegraphics[bb= 48 49 628 801,angle=0,width=16cm,clip]{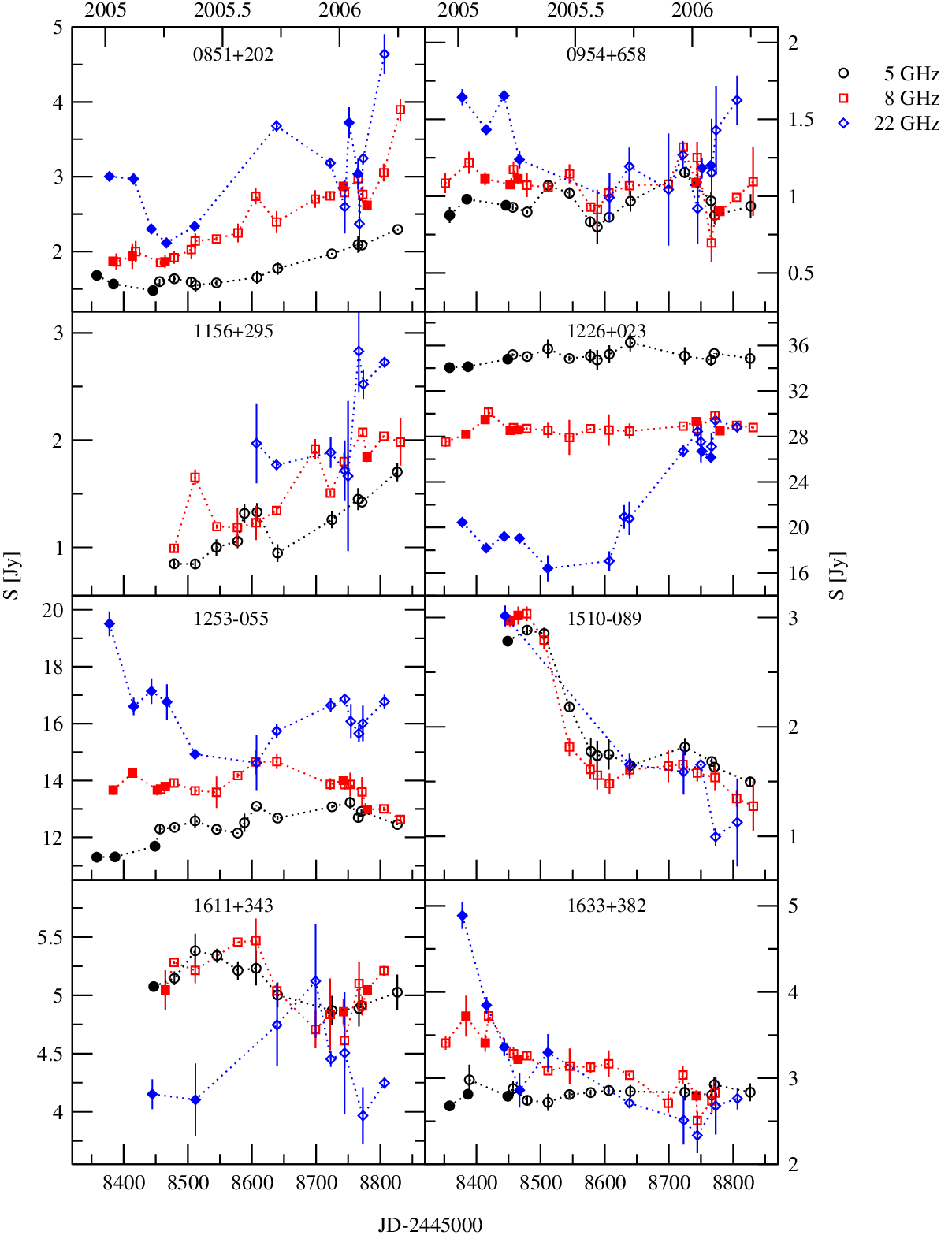}
\caption{continued.}
\end{figure*}
\setcounter{figure}{0}
\begin{figure*}[htbp]
\centering
\includegraphics[bb= 61 222 619 801,angle=0,width=16cm,clip]{6561_f1c.eps}
\caption{continued.}
\end{figure*}

\begin{figure*}[htbp]
\centering
\hbox{
\includegraphics[bb= 65 20 575 792,angle=-90,width=8cm,clip]{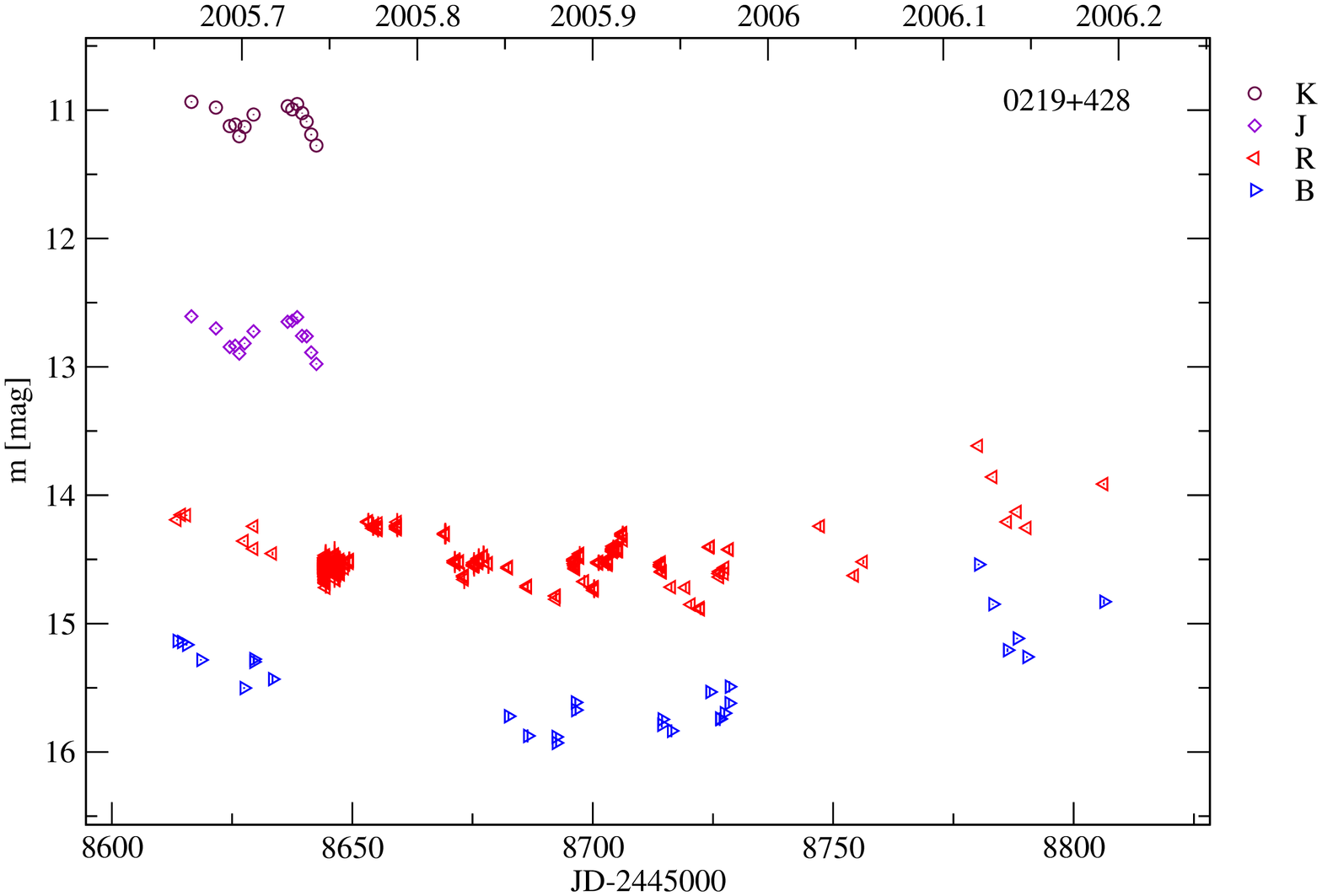}
\includegraphics[bb= 65 20 575 792,angle=-90,width=8cm,clip]{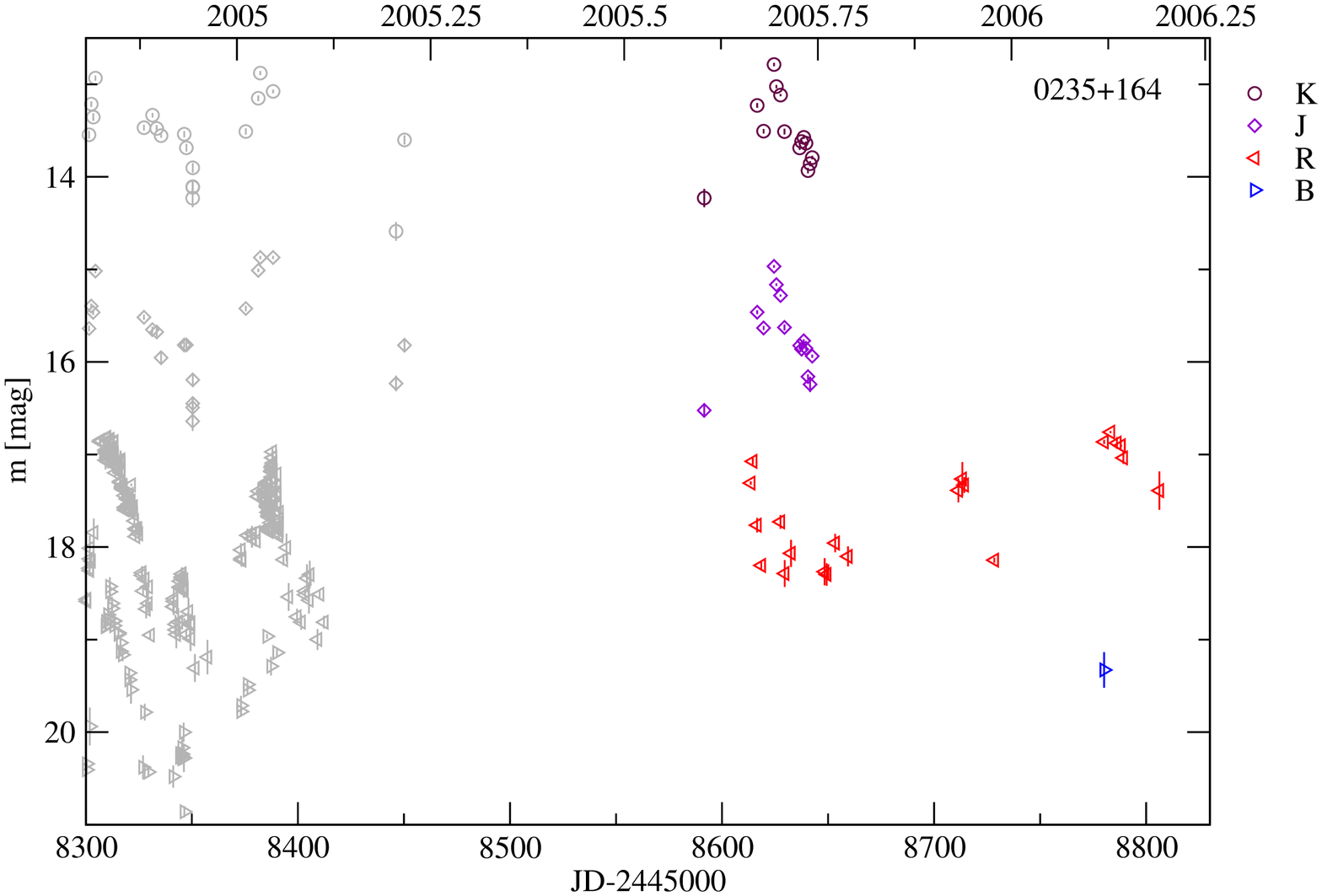}}
\hbox{
\includegraphics[bb= 65 20 575 792,angle=-90,width=8cm,clip]{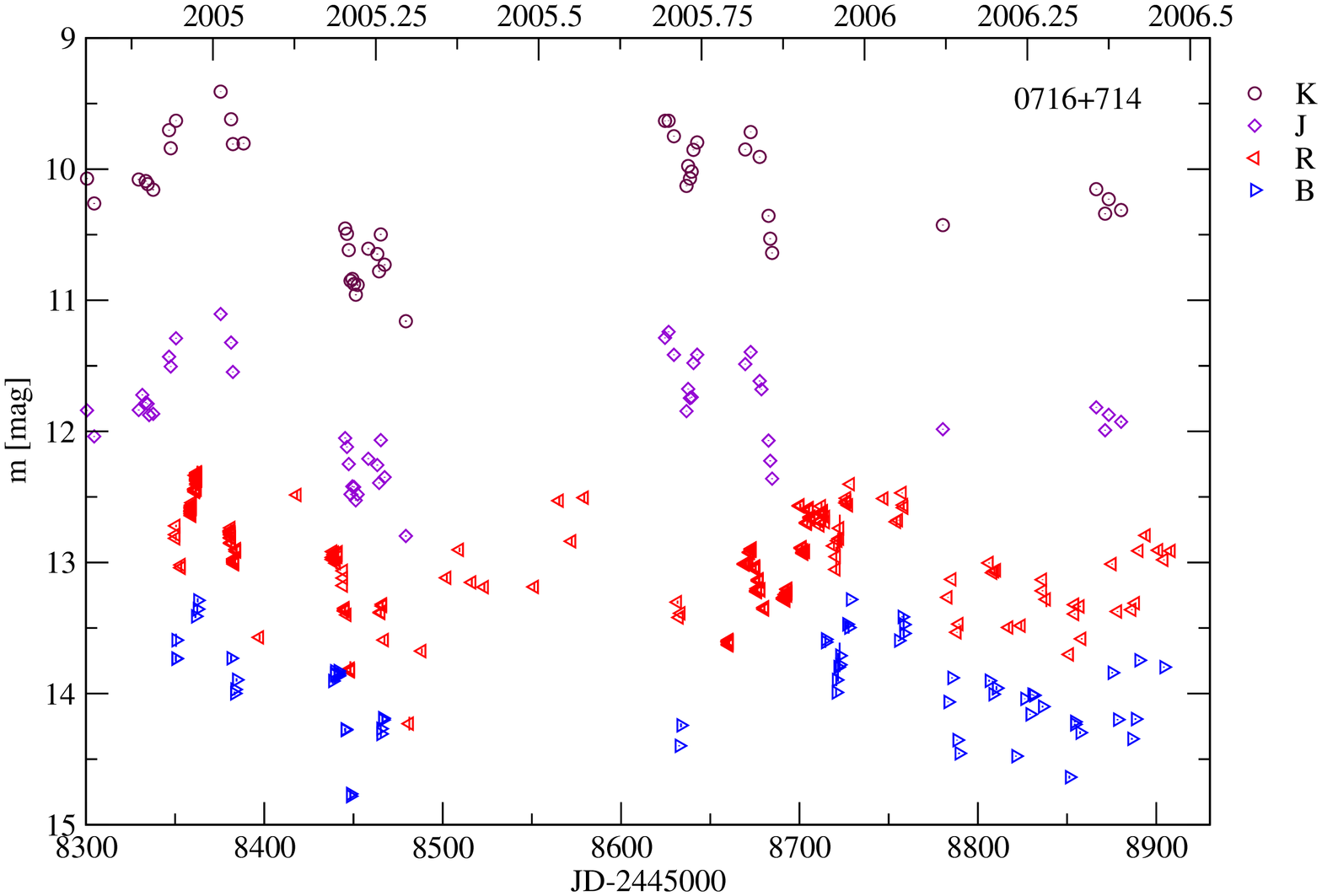}
\includegraphics[bb= 65 20 575 792,angle=-90,width=8cm,clip]{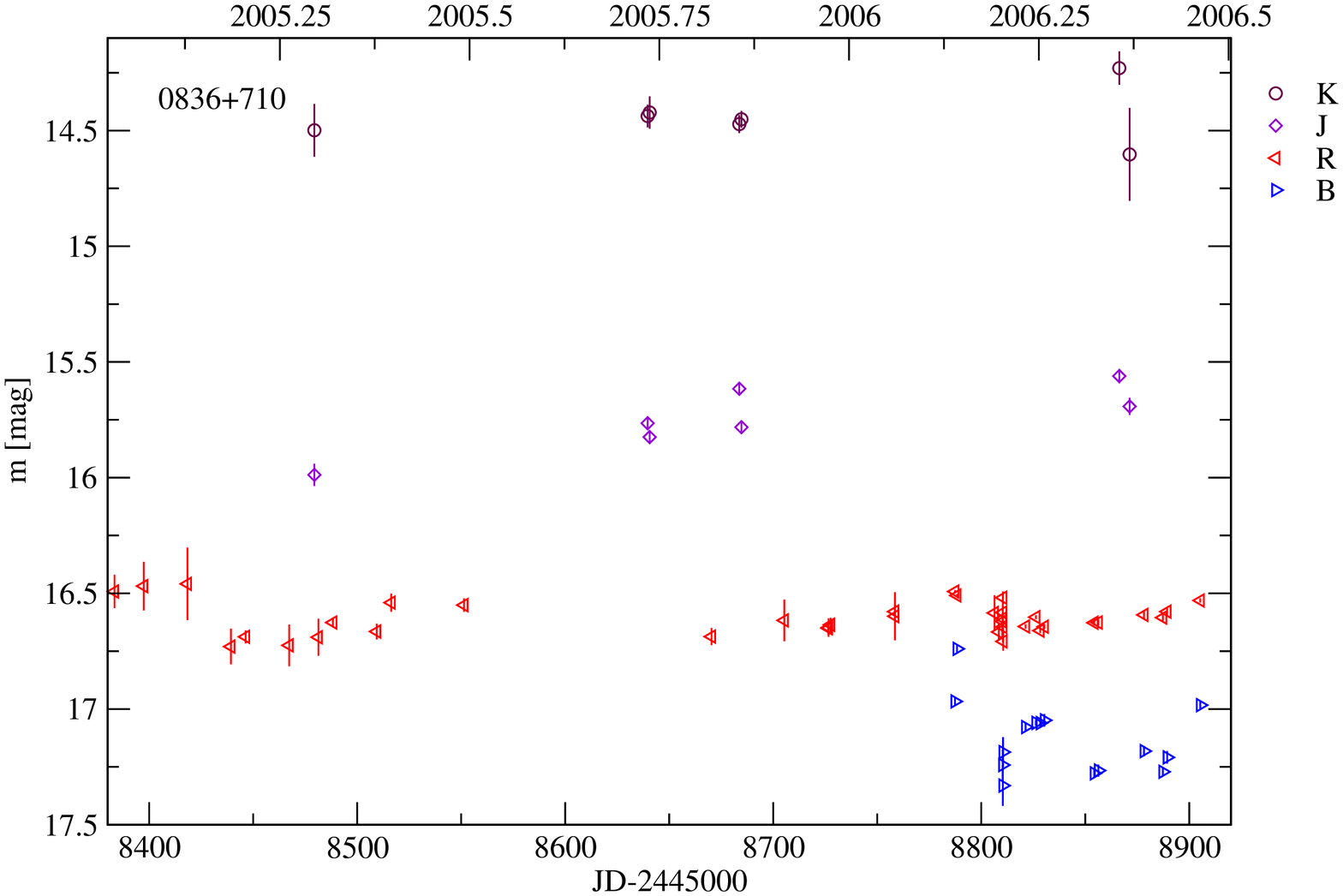}}
\hbox{
\includegraphics[bb= 65 20 575 792,angle=-90,width=8cm,clip]{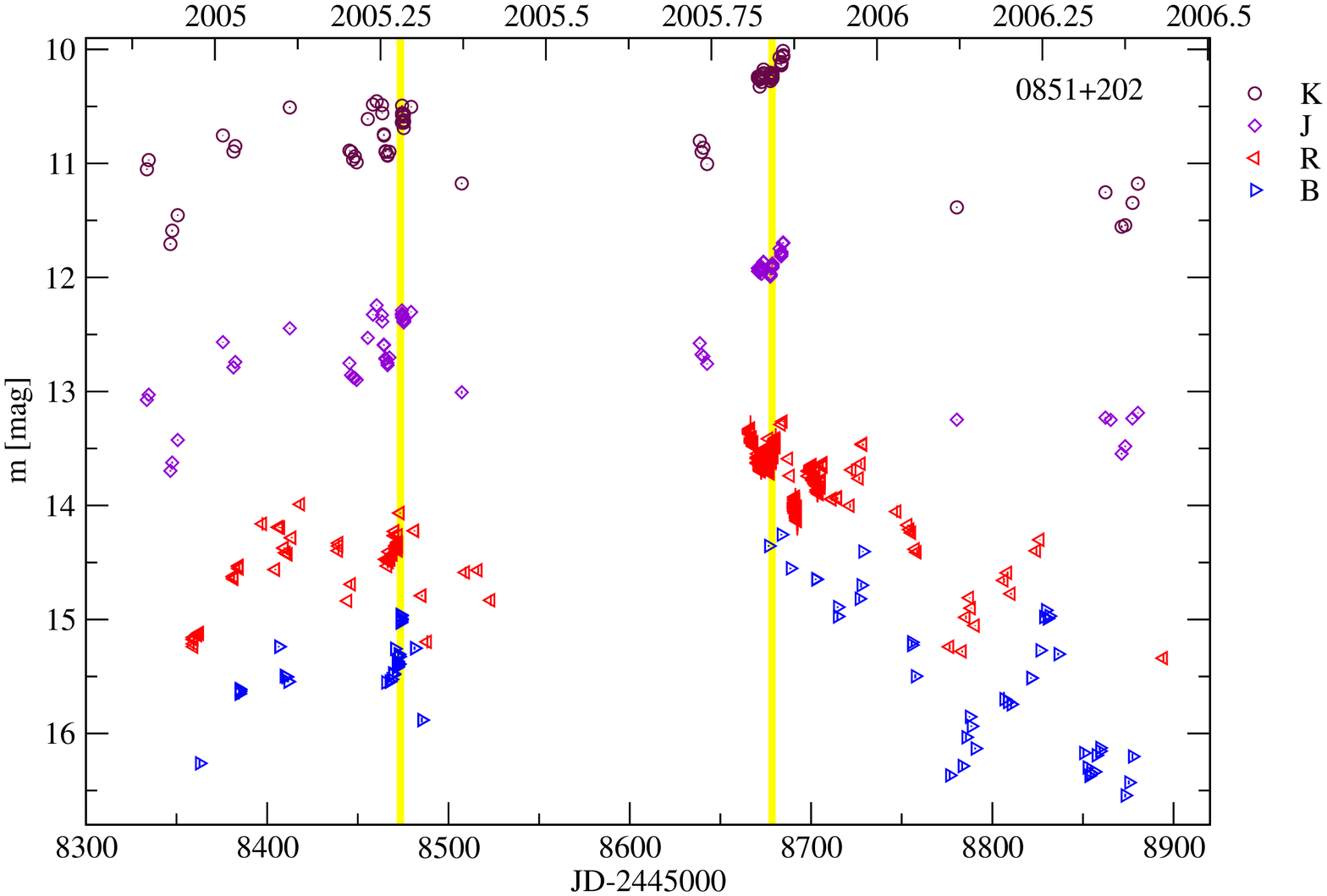}
\includegraphics[bb= 65 20 575 792,angle=-90,width=8cm,clip]{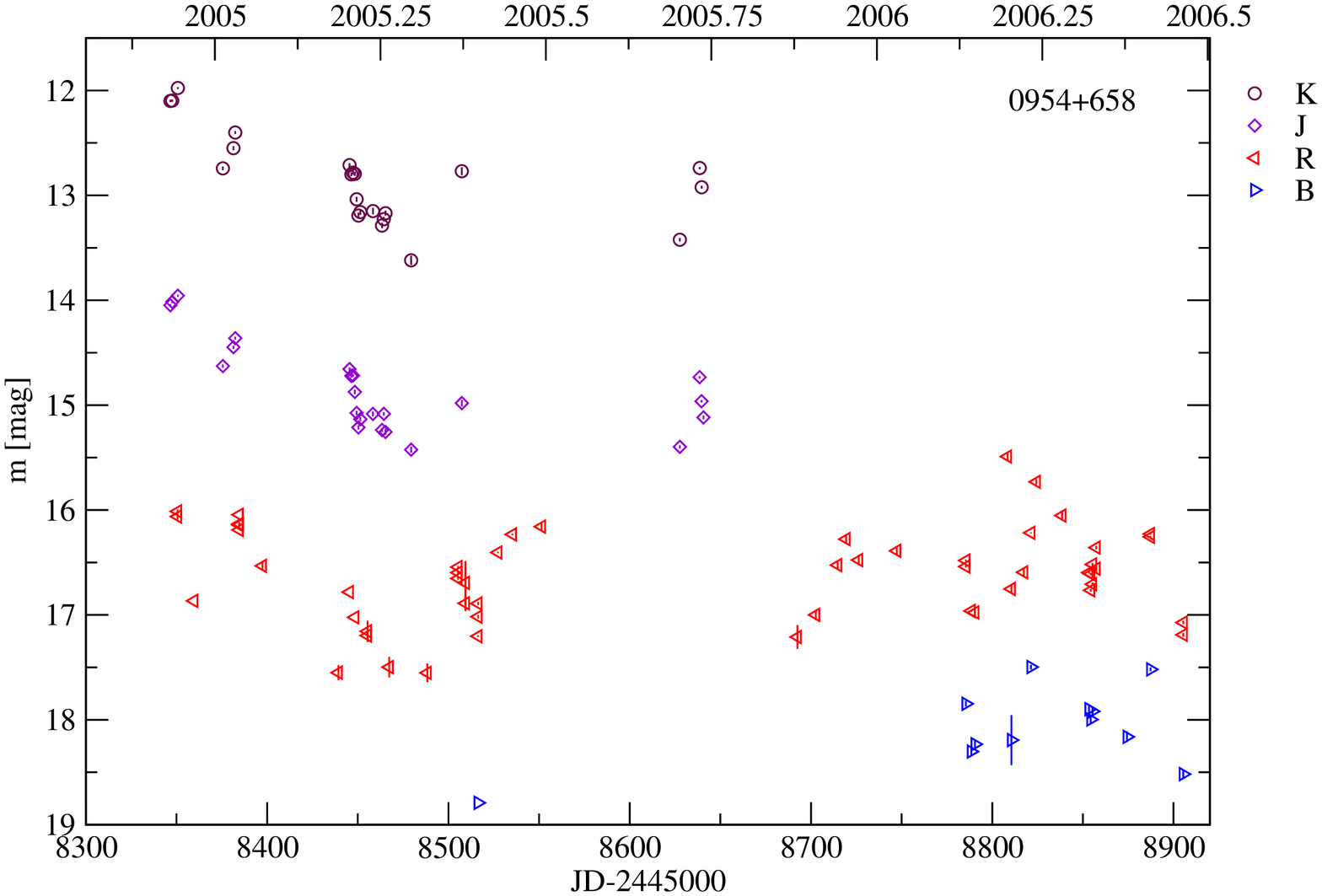}}
\hbox{
\includegraphics[bb= 65 20 575 792,angle=-90,width=8cm,clip]{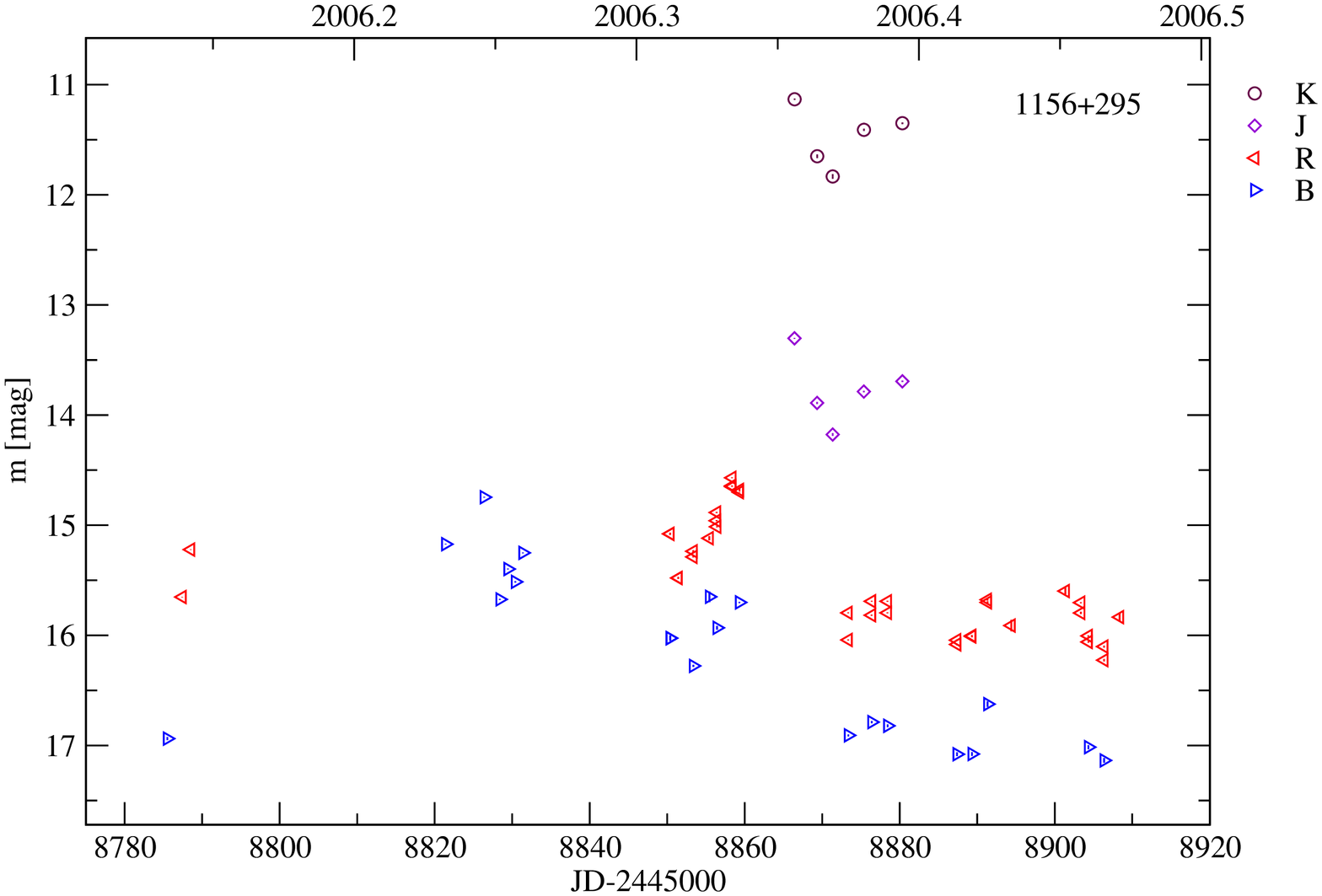}
\includegraphics[bb= 65 20 575 792,angle=-90,width=8cm,clip]{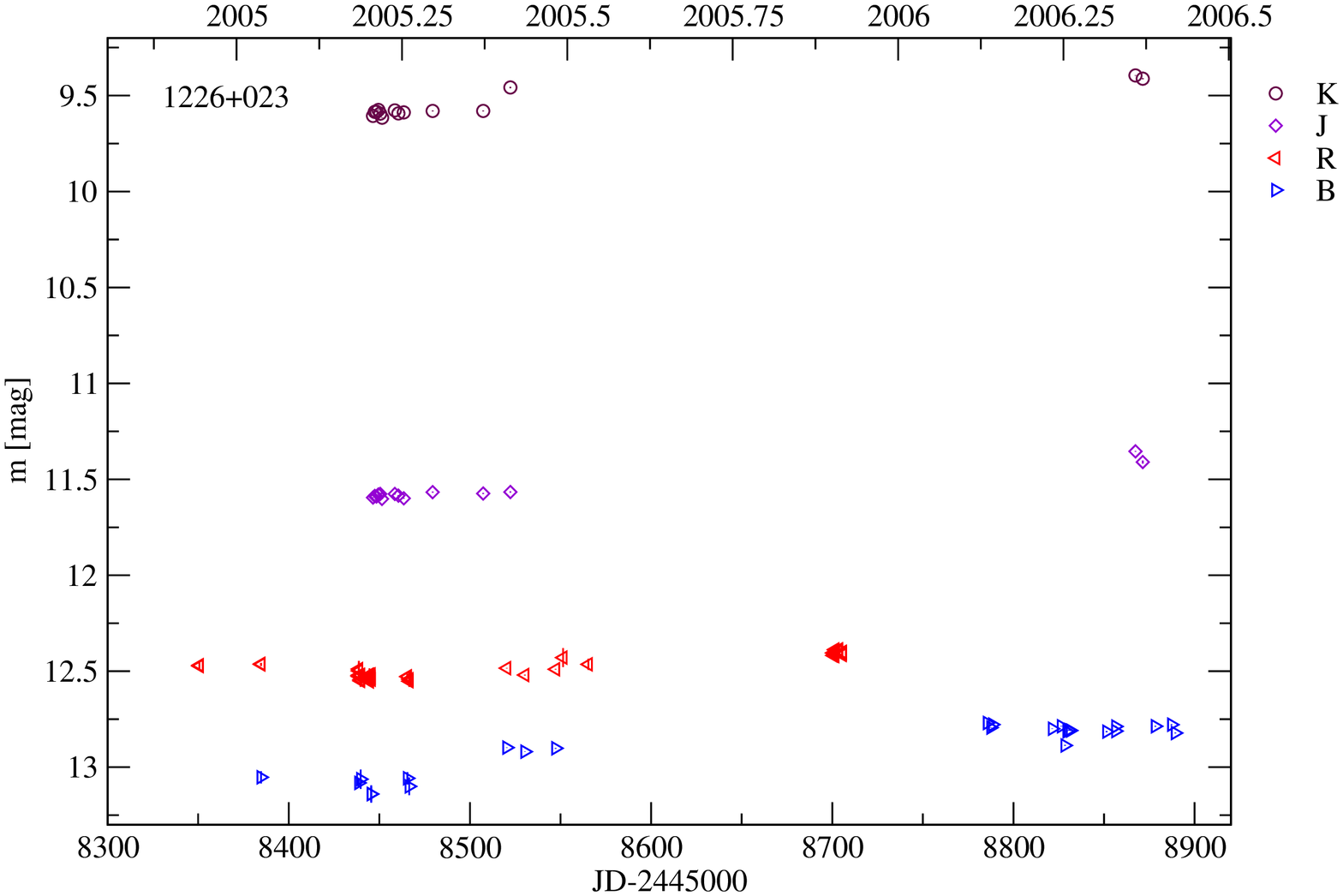}}
\caption{Near-IR  ($JK$) and optical ($BR$) light curves of those source where we
could get a good time coverage. The different colours (only in the electronic
version) and symbols denote the different observing bands. Grey symbols denote
already published data and yellow strips indicate the periods 
when WEBT campaigns were active, see Sect.\ref{sec:olc}
for more details. Note that the time range shown is not
the same for all sources.} \label{fig:olc} 
\end{figure*}

\setcounter{figure}{1}
\begin{figure*}[htbp]
\centering
\hbox{
\includegraphics[bb= 65 20 575 792,angle=-90,width=9cm,clip]{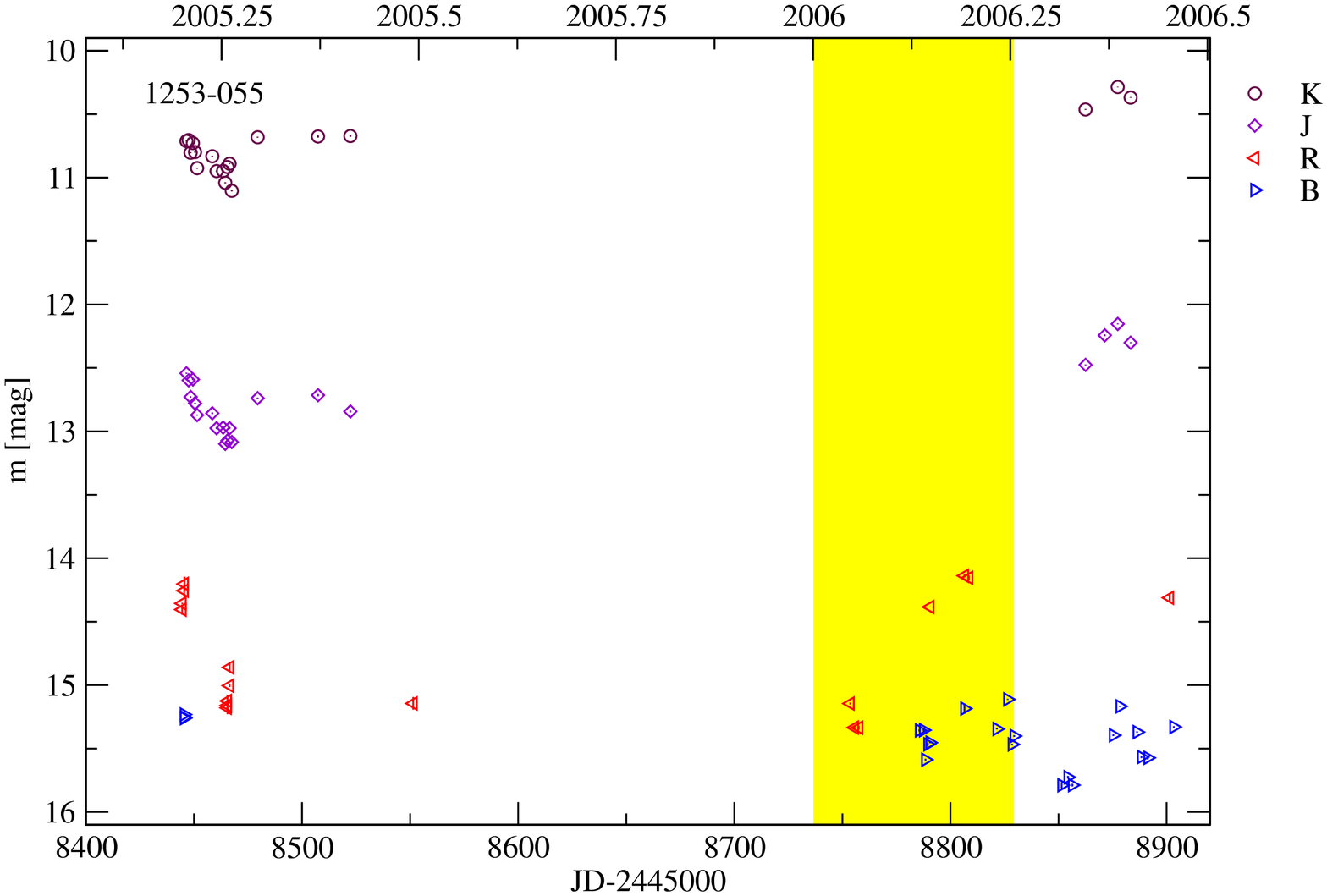}
\includegraphics[bb= 65 20 575 792,angle=-90,width=9cm,clip]{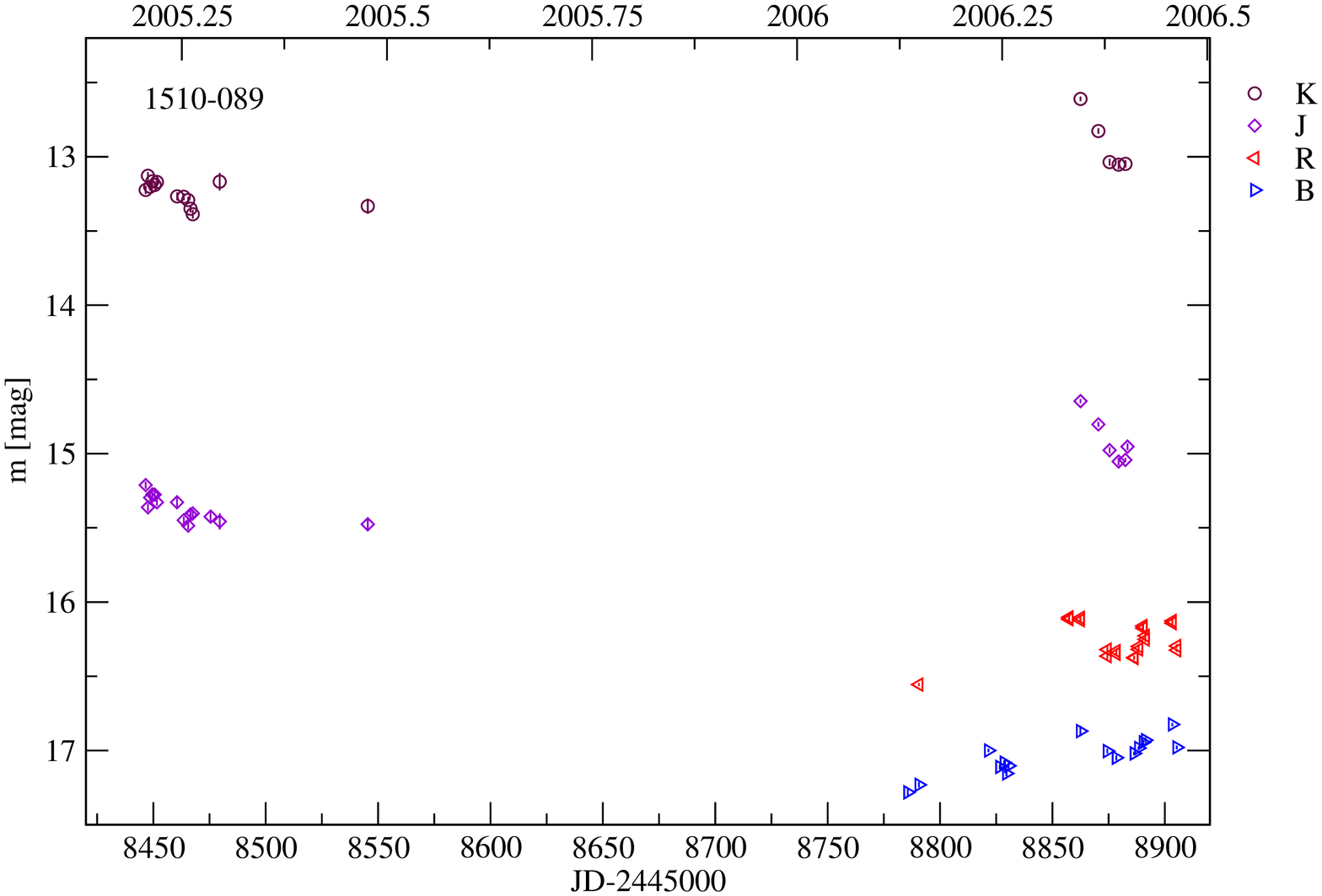}}
\hbox{
\includegraphics[bb= 65 20 575 792,angle=-90,width=9cm,clip]{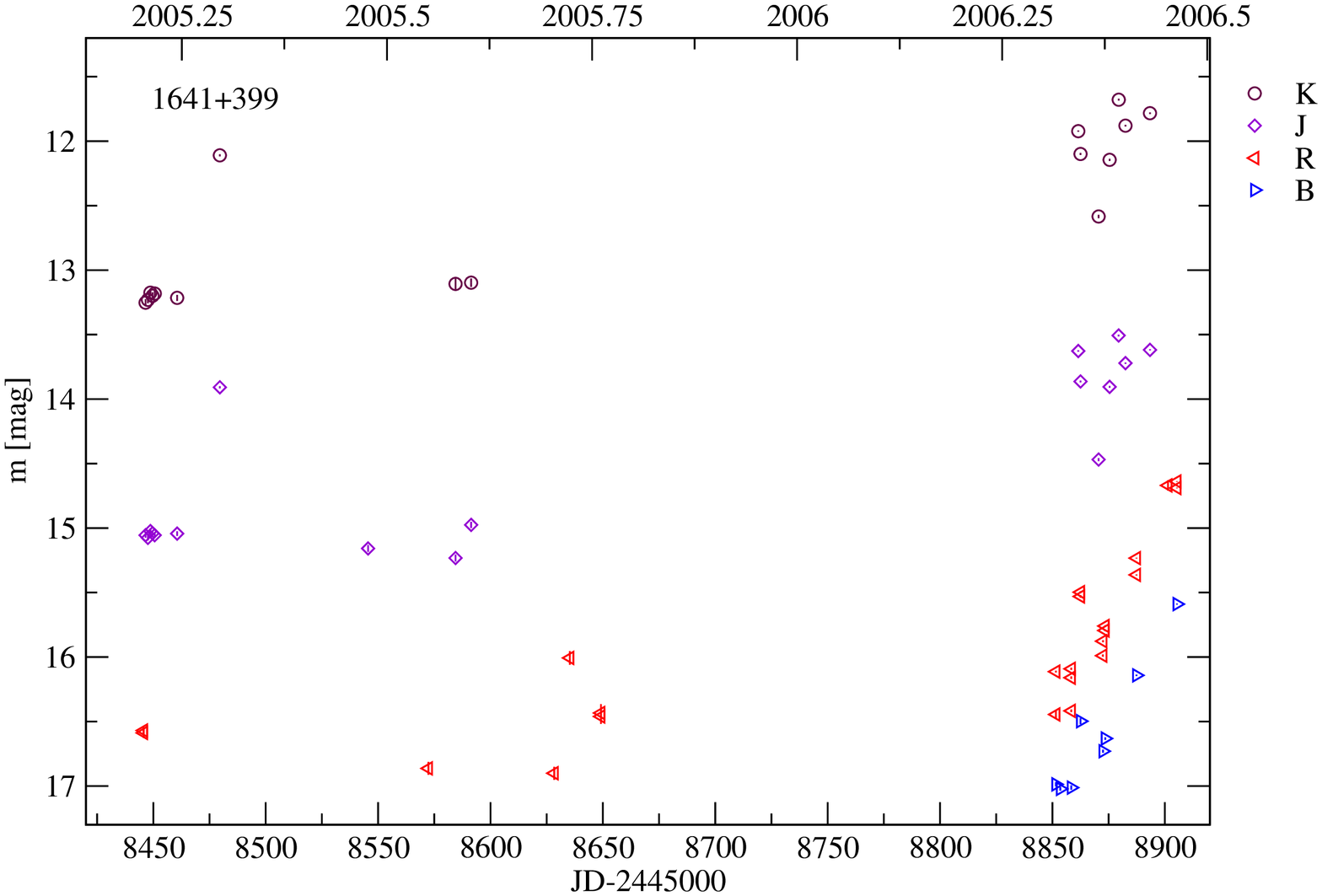}
\includegraphics[bb= 65 20 575 792,angle=-90,width=9cm,clip]{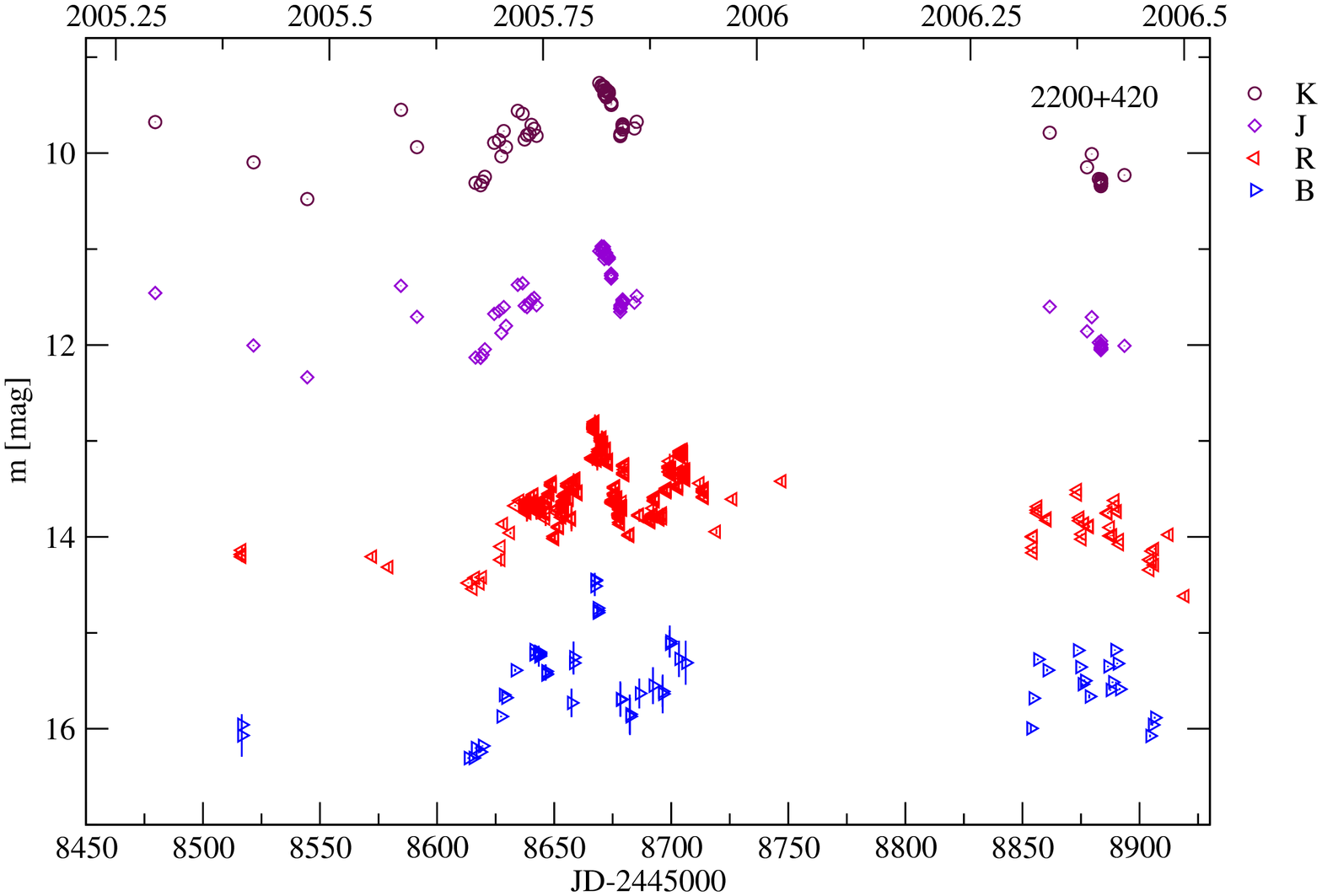}}
\hbox{
\includegraphics[bb= 65 20 575
792,angle=-90,width=9cm,clip]{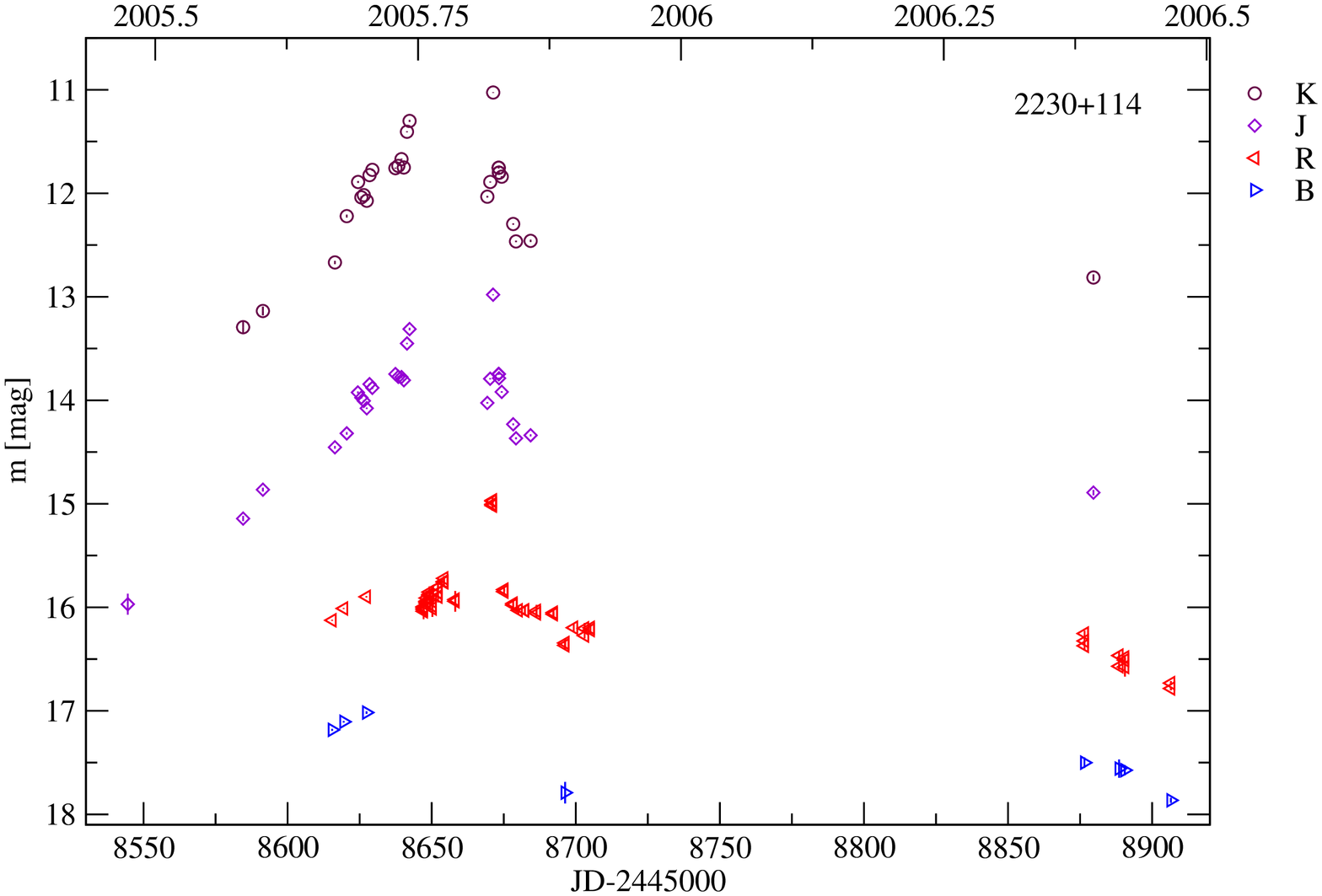}
\includegraphics[bb= 65 20 575
792,angle=-90,width=9cm,clip]{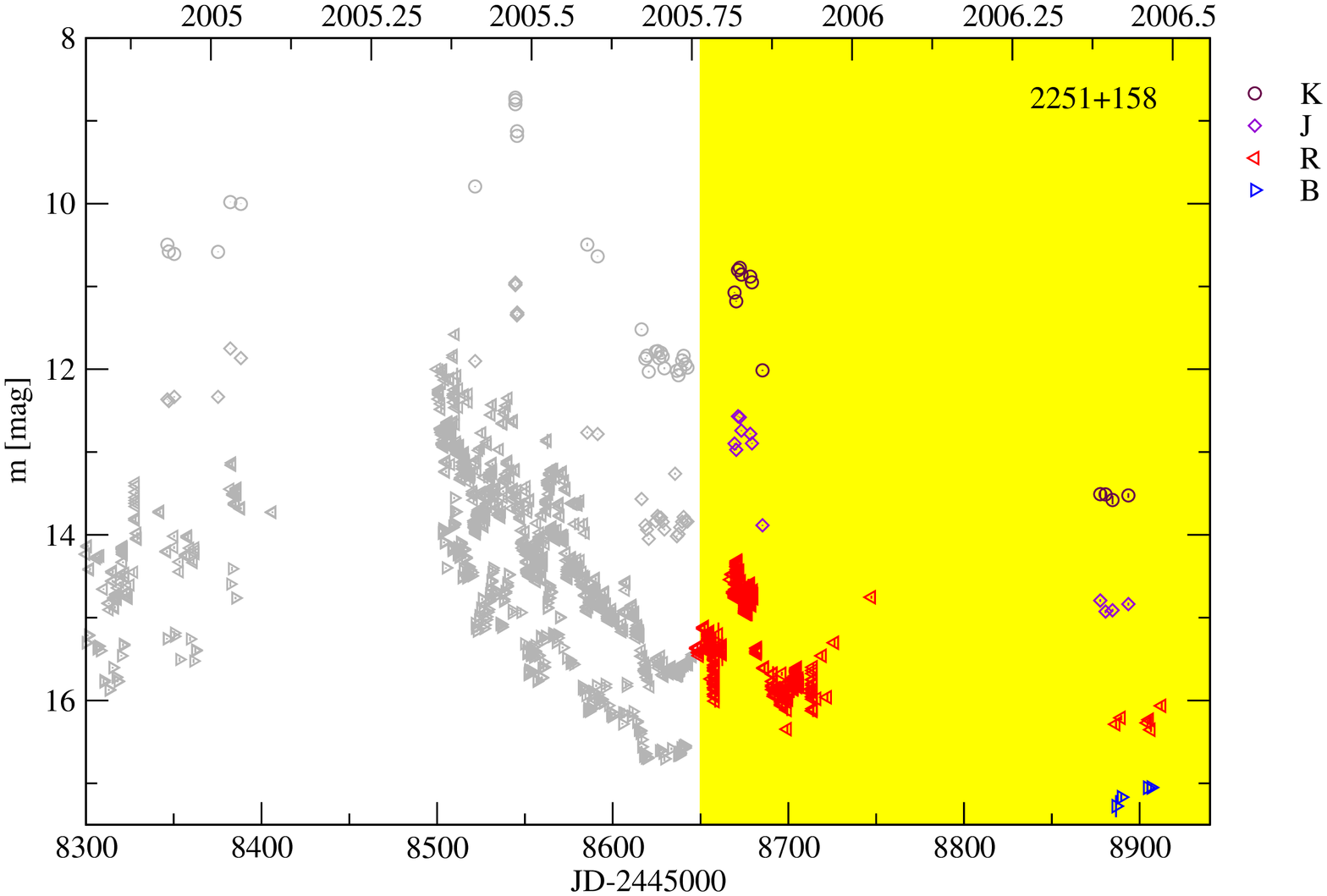}}
\caption{continued.}\label{fig:olc2}
\end{figure*}

\begin{figure*}[htbp]
\centering
\hbox{
\includegraphics[bb= 40 0 575 842,angle=-90,width=9cm,clip]{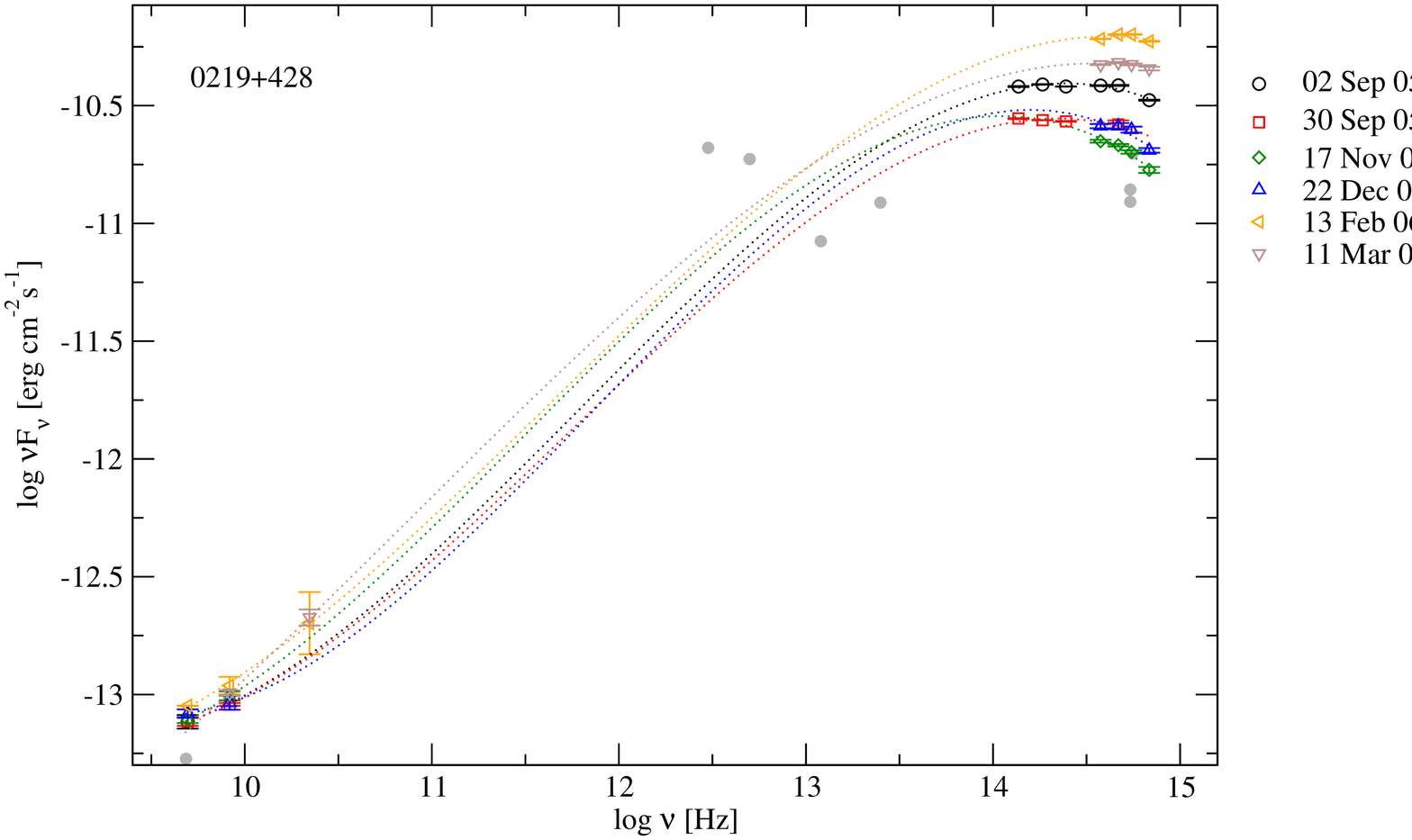}
\includegraphics[bb= 40 0 575 842,angle=-90,width=9cm,clip]{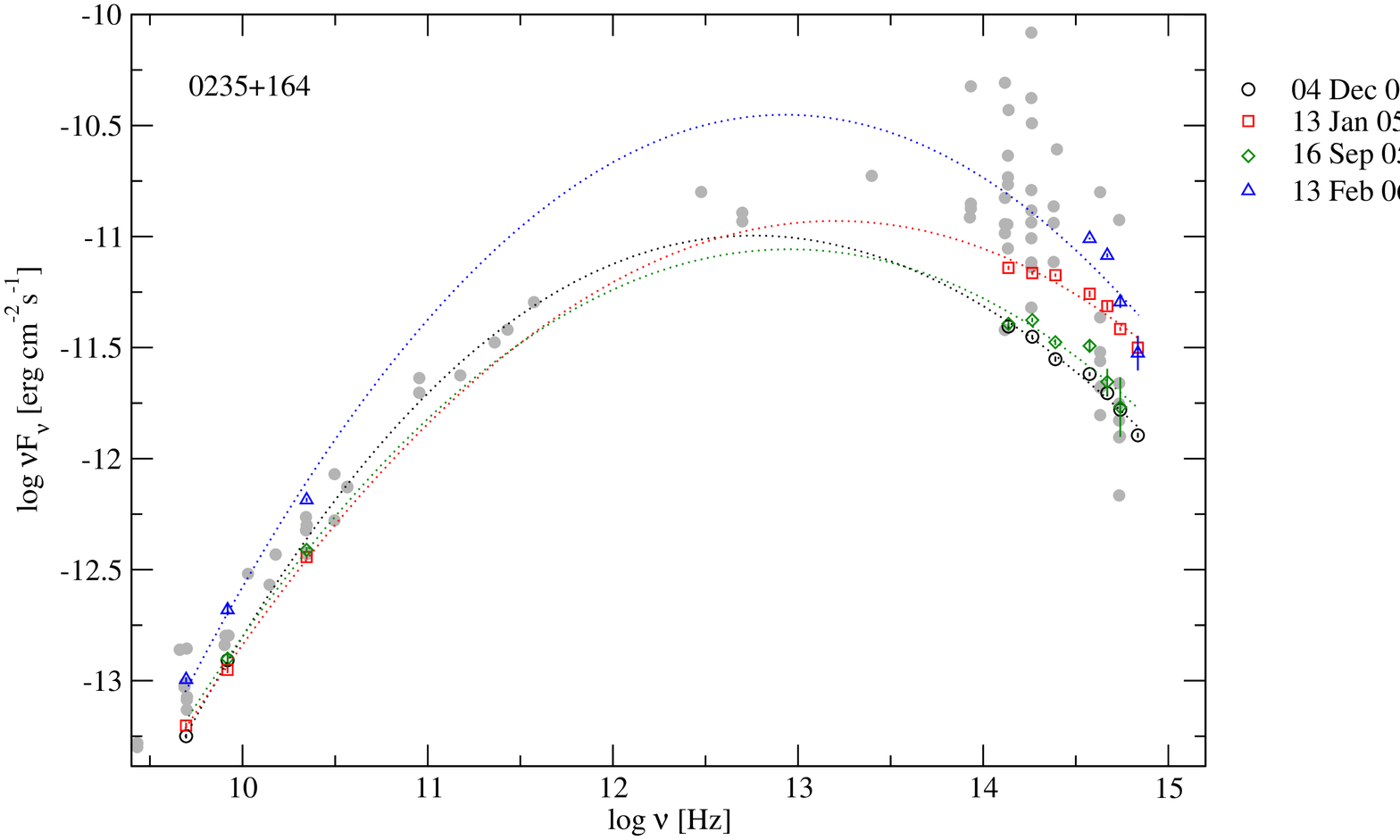}}
\hbox{
\includegraphics[bb= 40 0 575 842,angle=-90,width=9cm,clip]{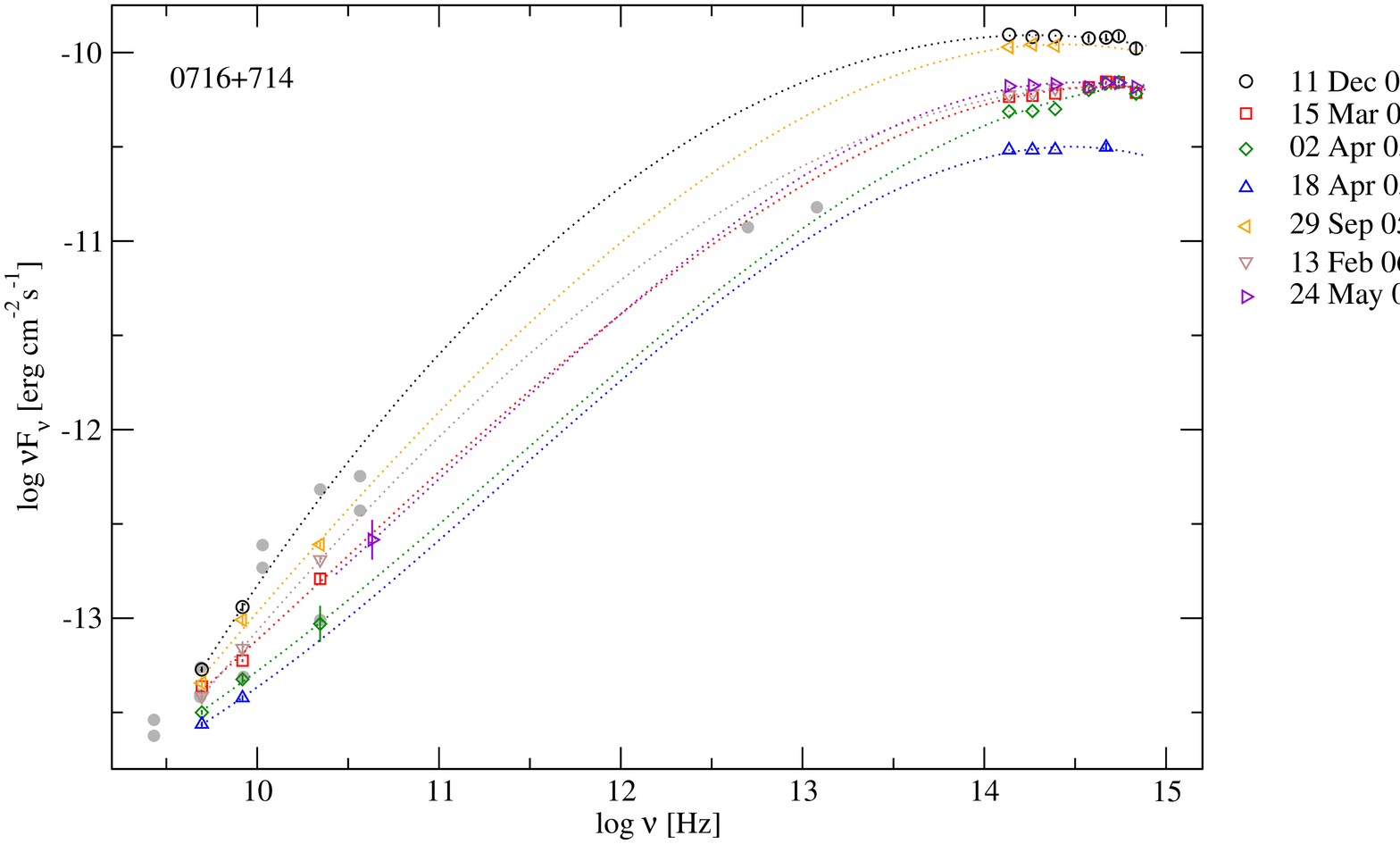}
\includegraphics[bb= 40 0 575 842,angle=-90,width=9cm,clip]{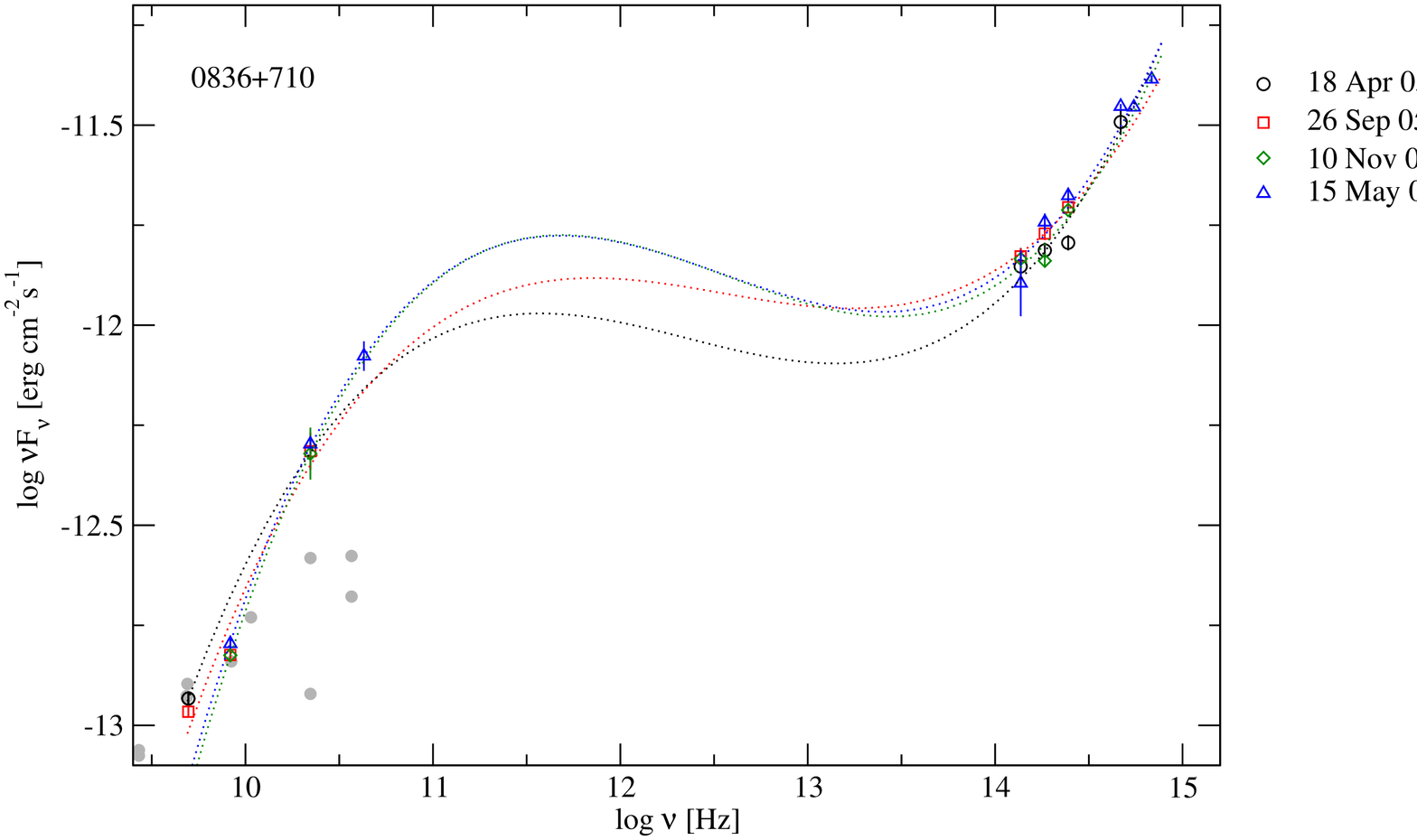}}
\hbox{
\includegraphics[bb= 40 0 575 842,angle=-90,width=9cm,clip]{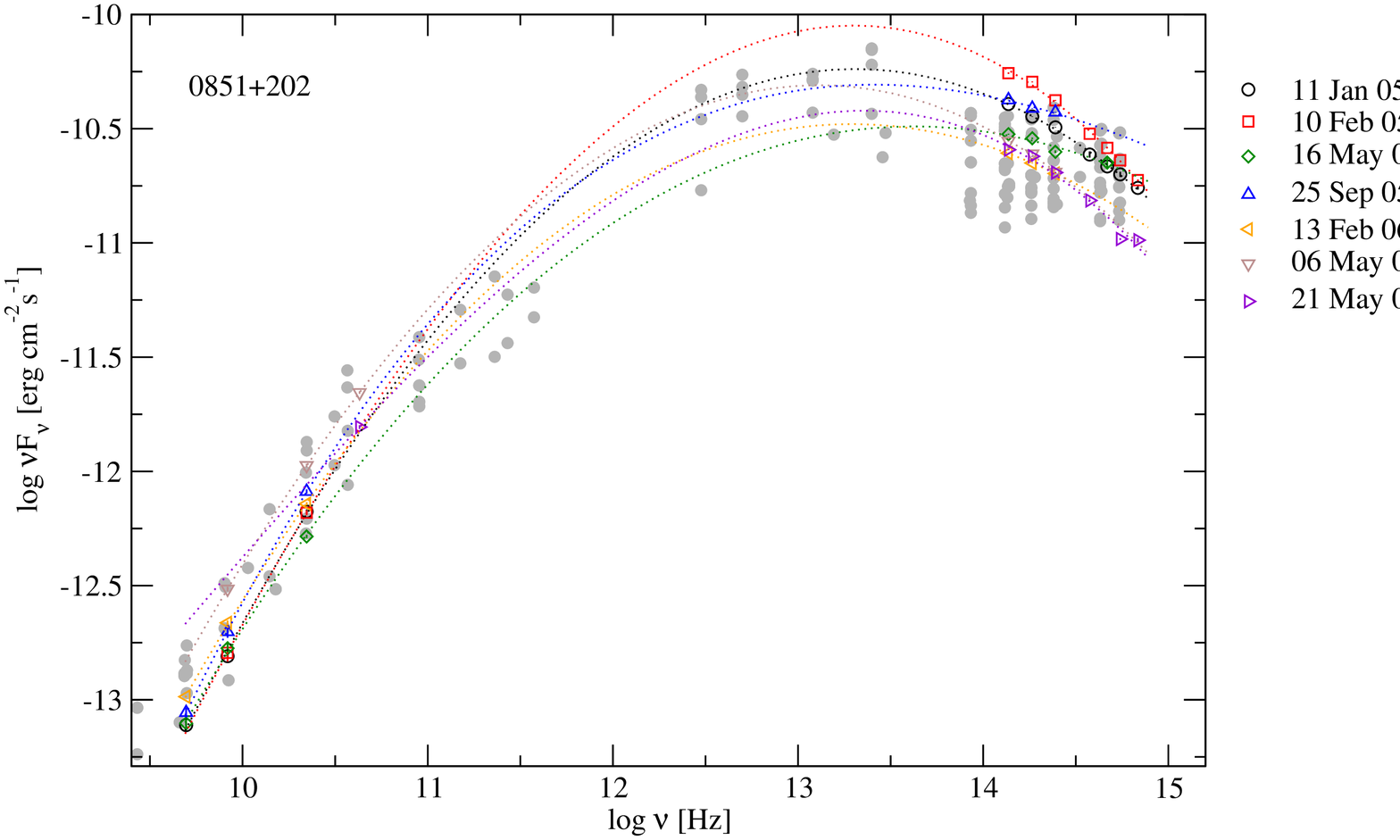}
\includegraphics[bb= 40 0 575 842,angle=-90,width=9cm,clip]{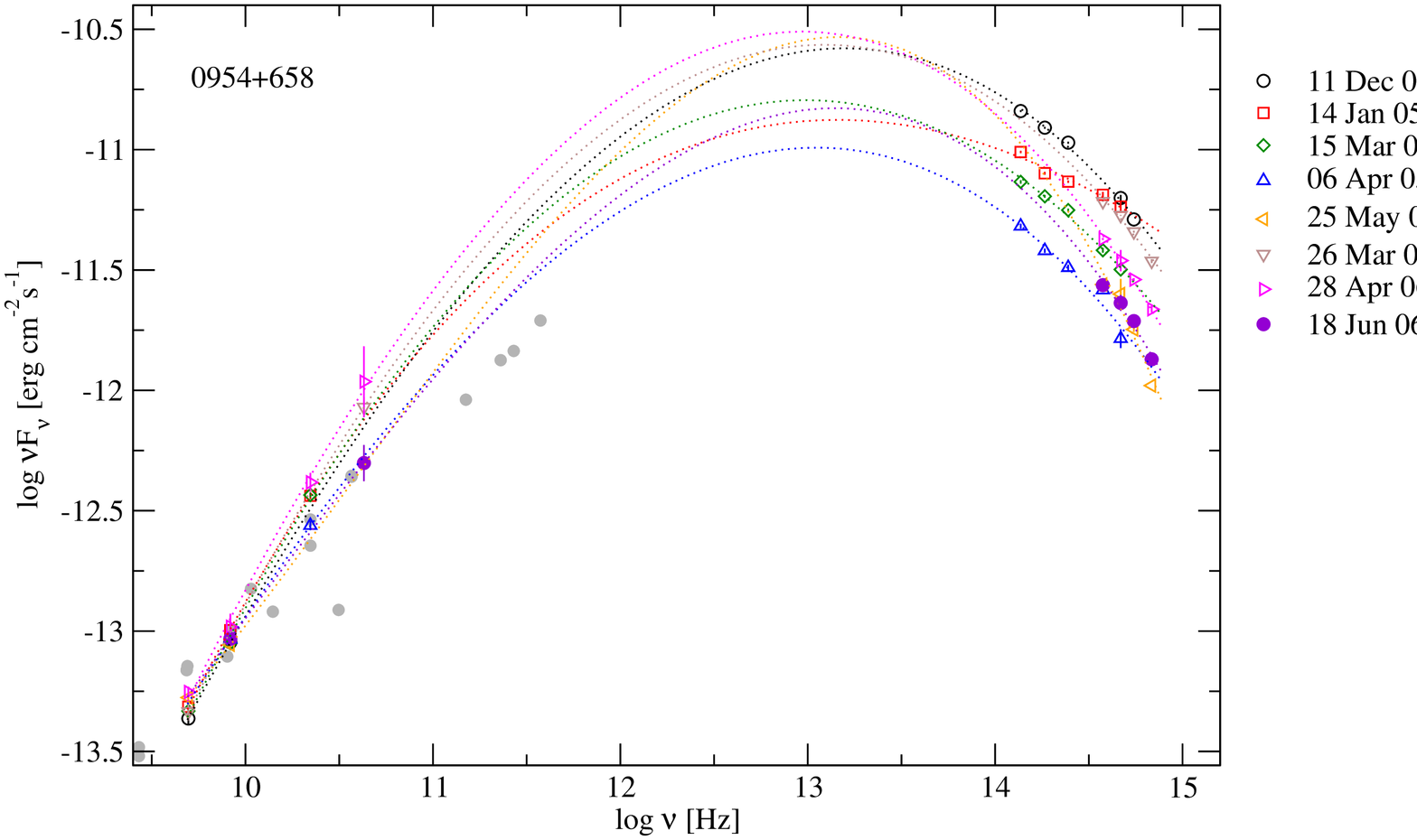}}
\hbox{
\includegraphics[bb= 40 0 575 842,angle=-90,width=9cm,clip]{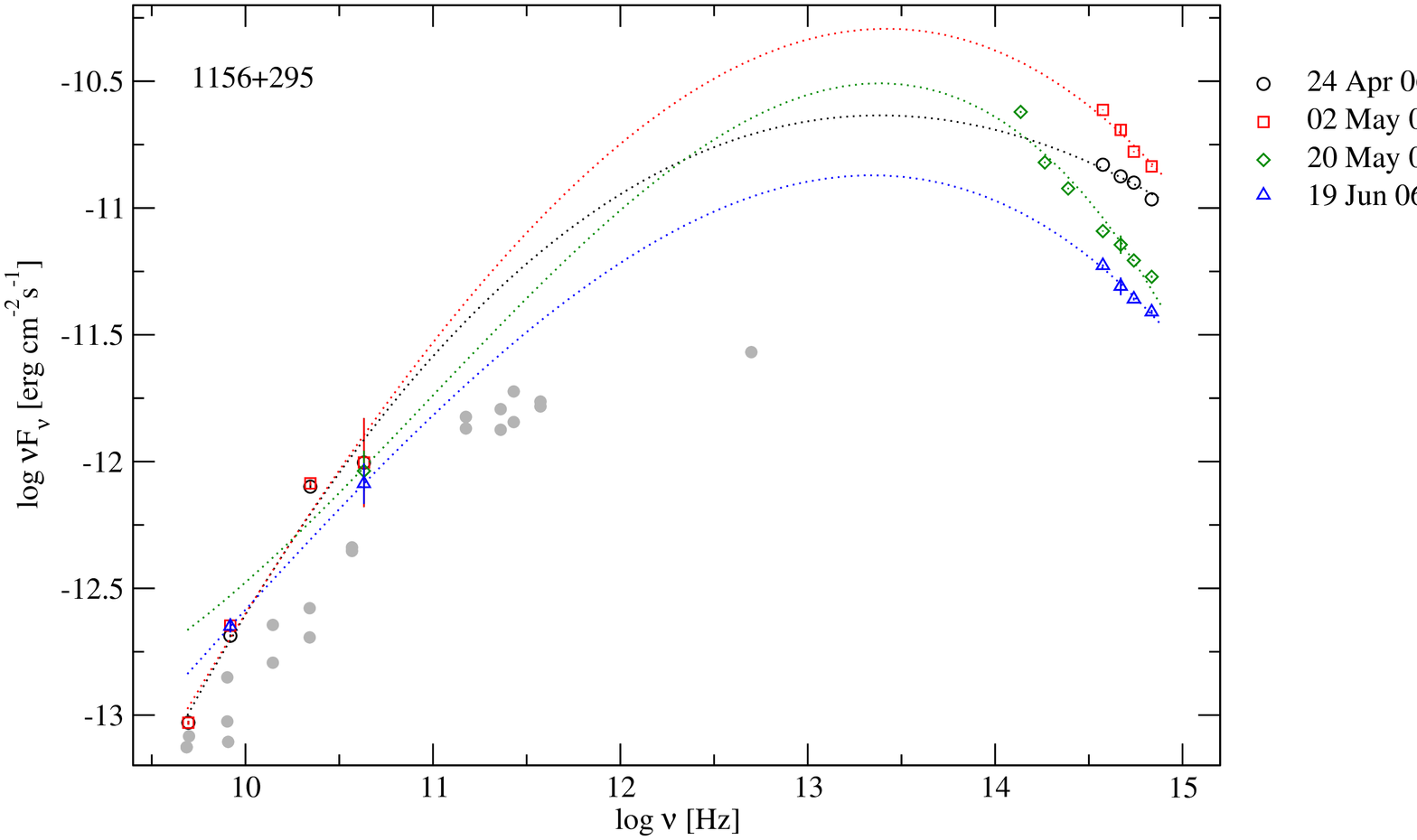}
\includegraphics[bb= 40 0 575 842,angle=-90,width=9cm,clip]{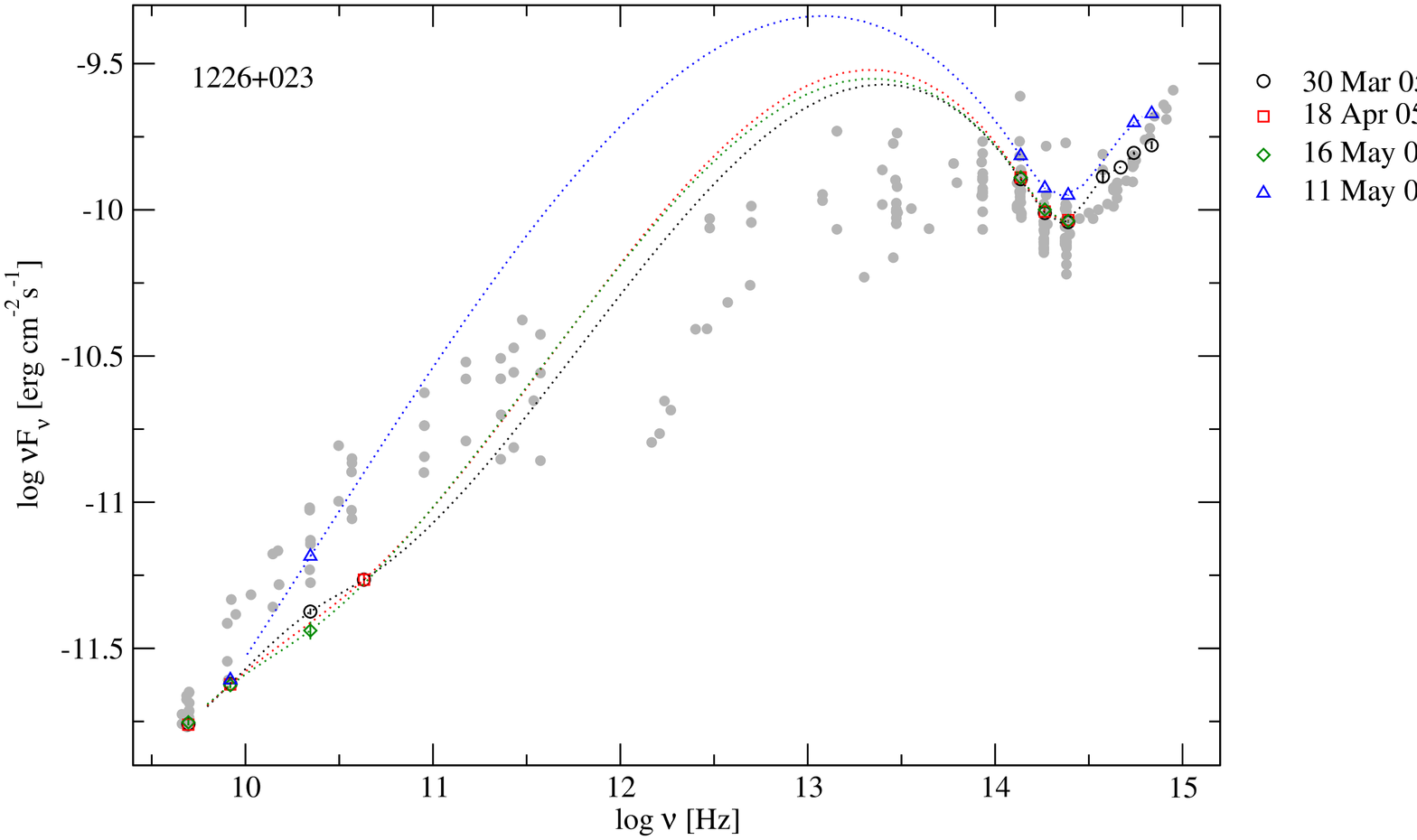}}
\caption{Examples of blazar SEDs, which we obtain from our data for those
sources which are best sampled. Grey dots represent archival data taken from NED.
See Sect.~\ref{sec:individ} for more details.}\label{fig:SEDs}
\end{figure*}

\setcounter{figure}{2}
\begin{figure*}[htbp]
\centering
\hbox{
\includegraphics[bb= 40 0 575 842,angle=-90,width=9cm,clip]{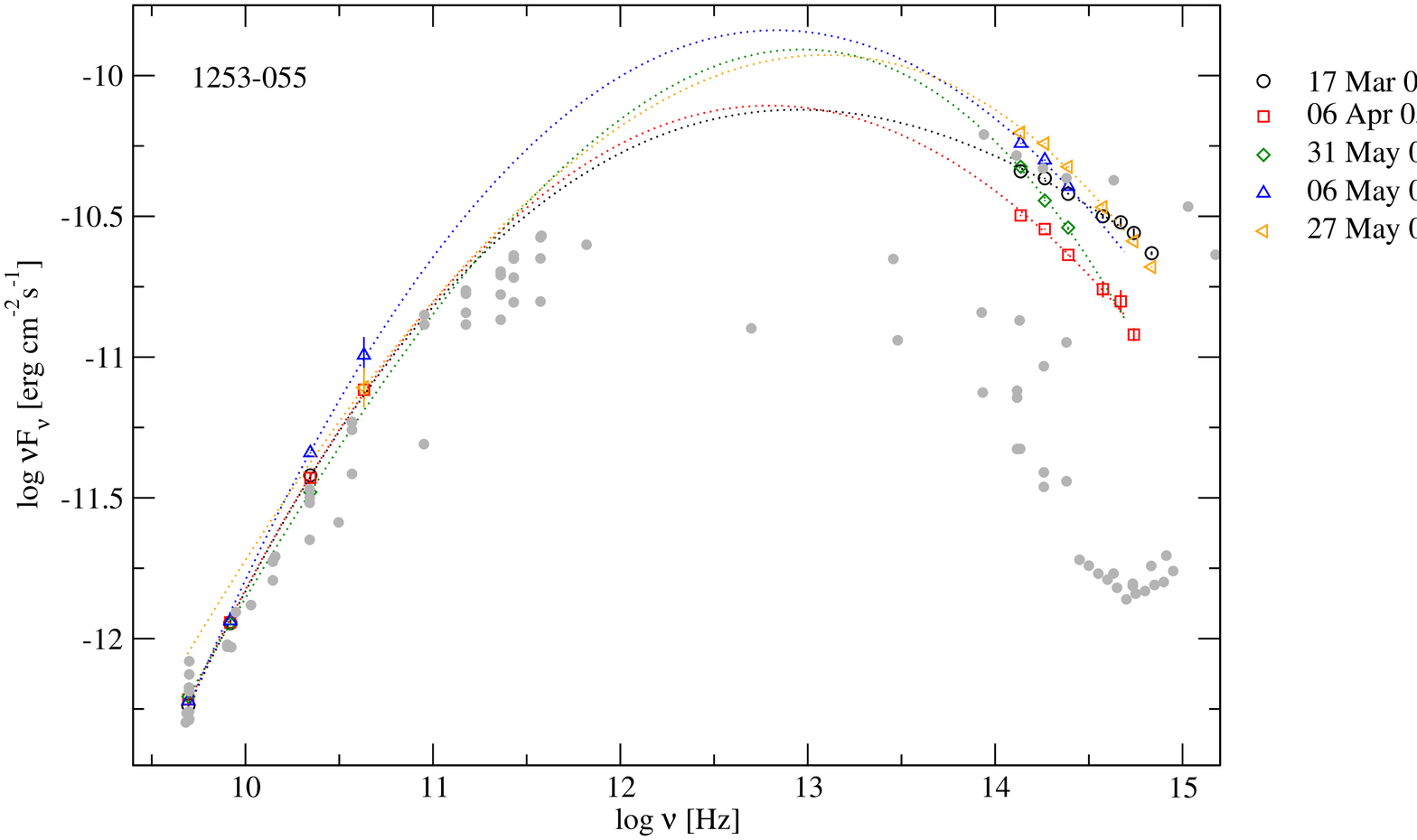}
\includegraphics[bb= 40 0 575 842,angle=-90,width=9cm,clip]{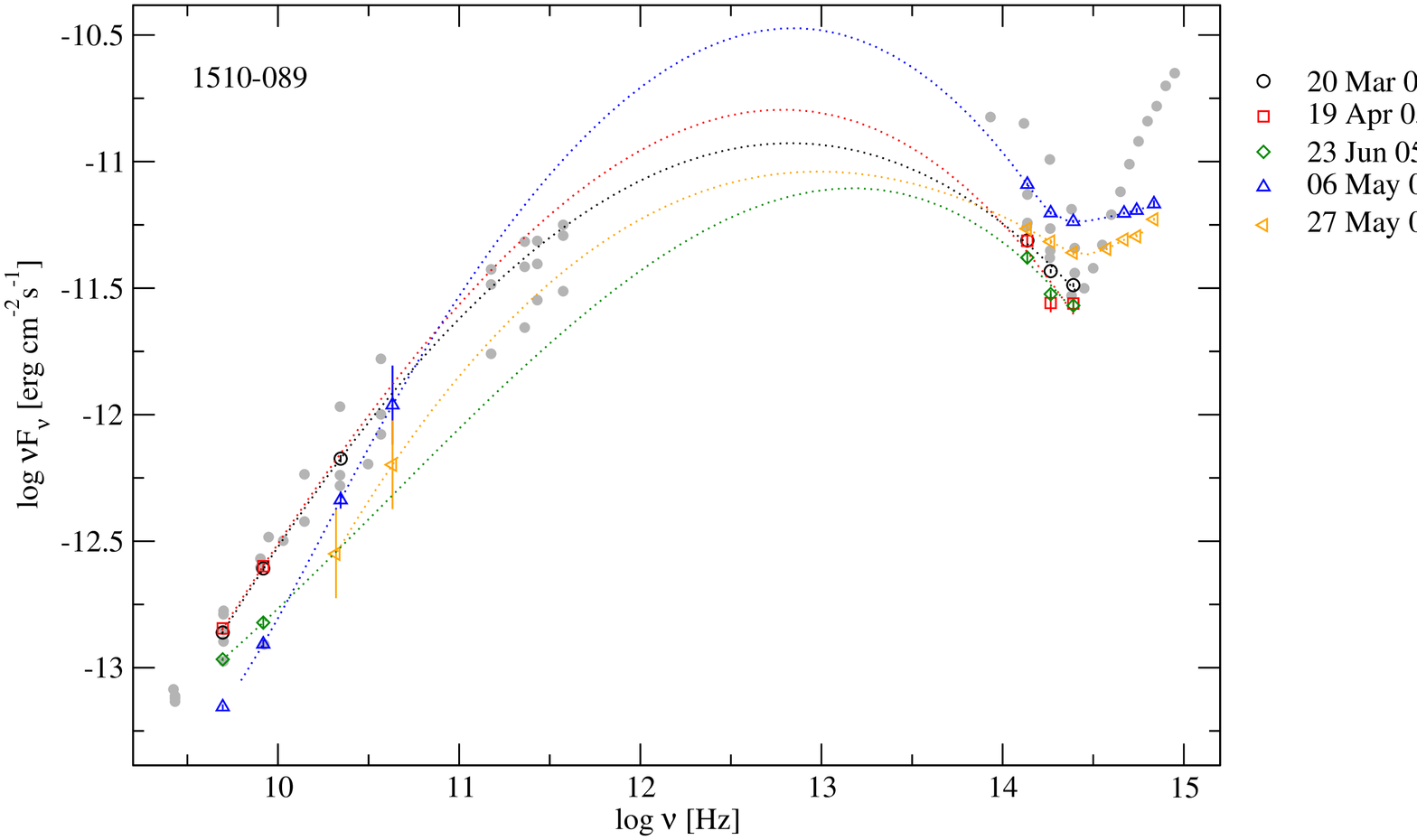}}
\hbox{
\includegraphics[bb= 40 0 575 842,angle=-90,width=9cm,clip]{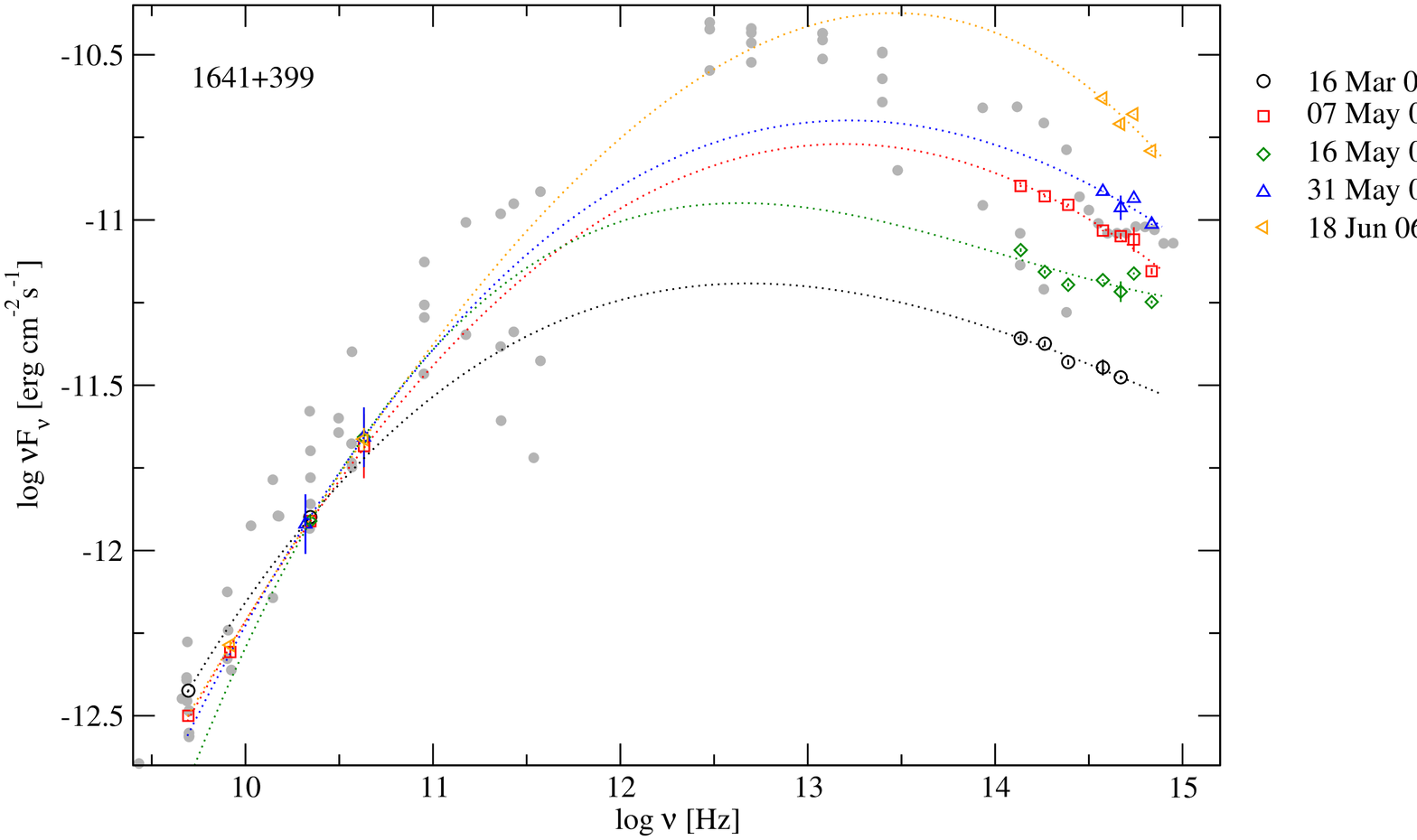}
\includegraphics[bb= 40 0 575 842,angle=-90,width=9cm,clip]{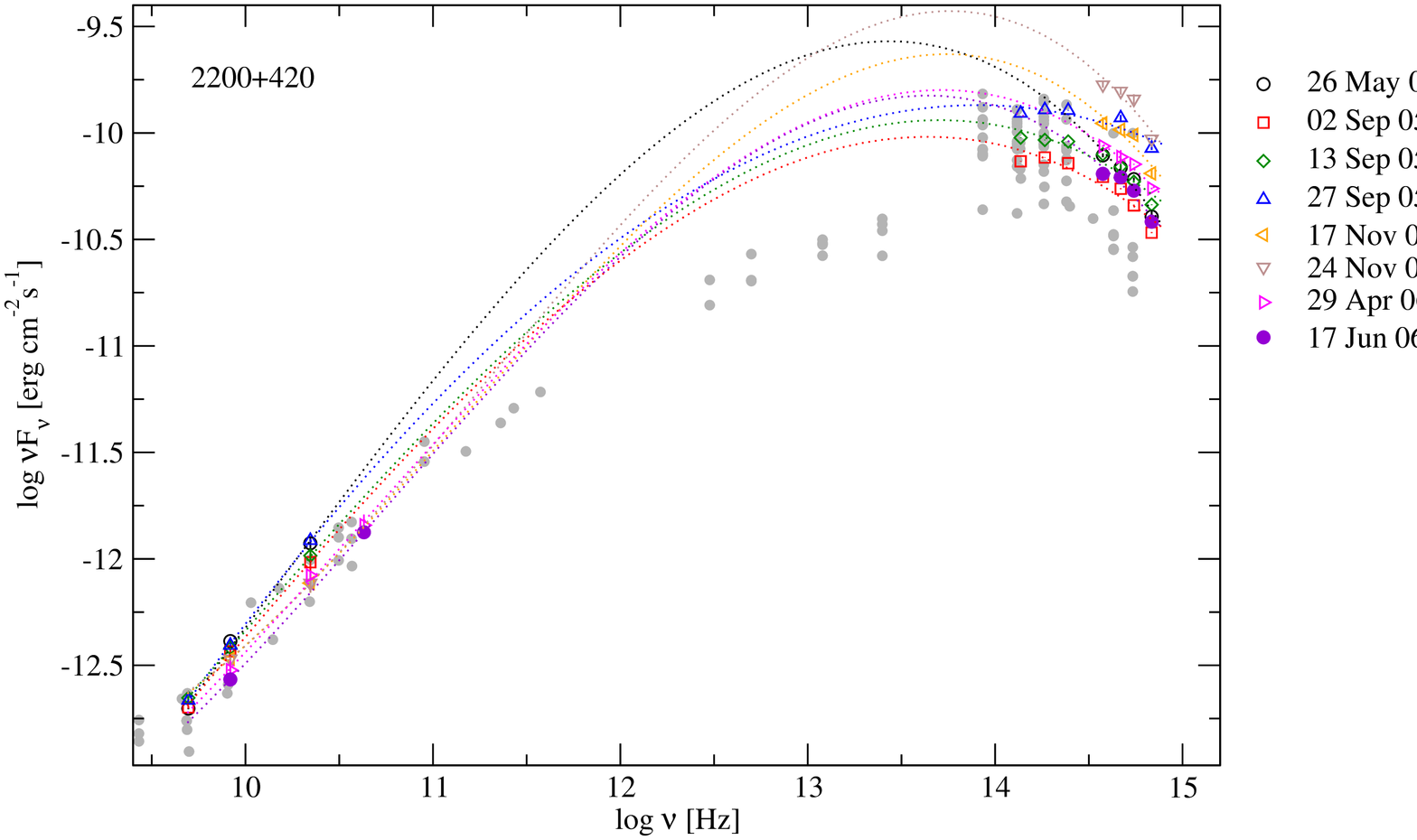}}
\hbox{
\includegraphics[bb= 40 0 575 842,angle=-90,width=9cm,clip]{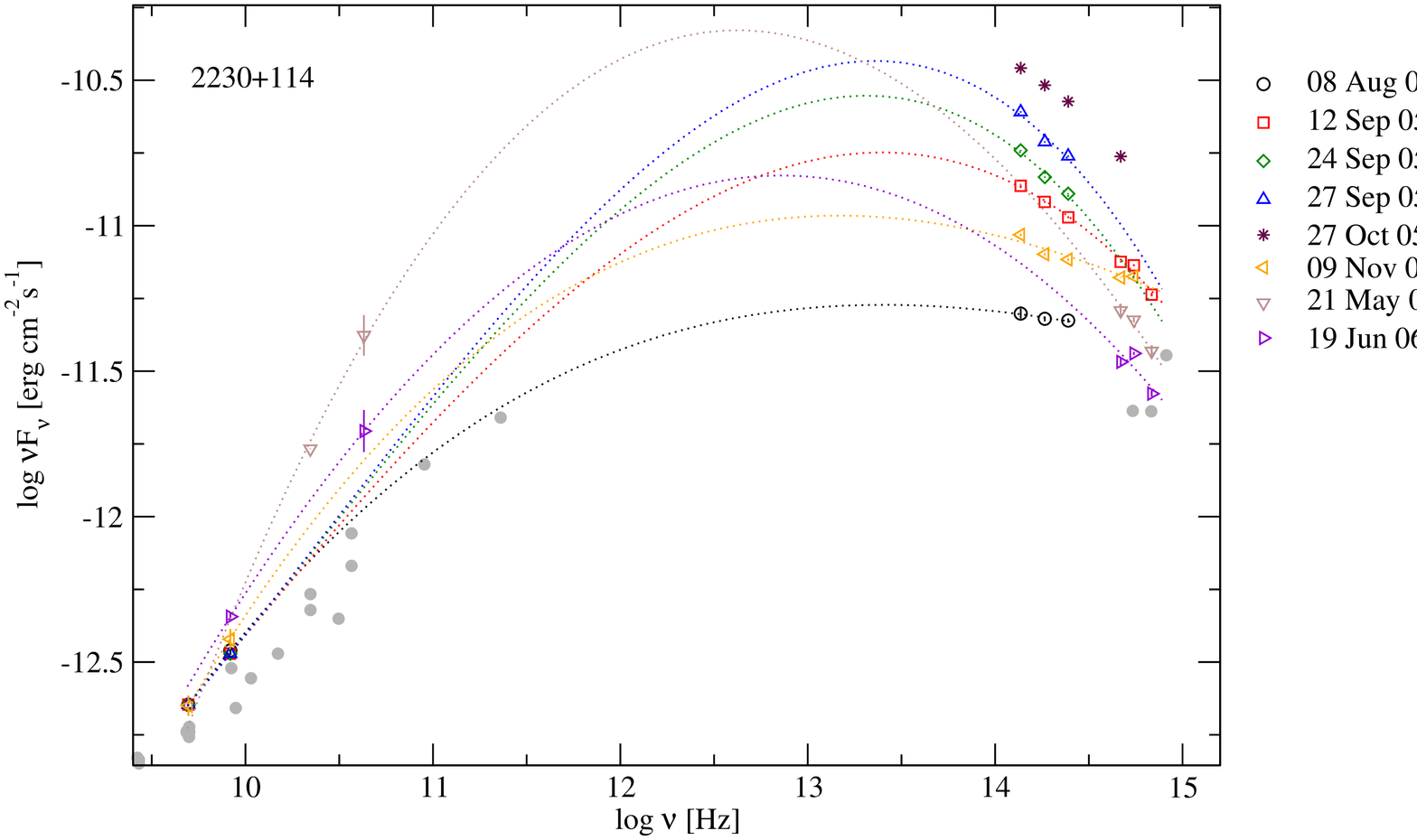}
\includegraphics[bb= 40 0 575 842,angle=-90,width=9cm,clip]{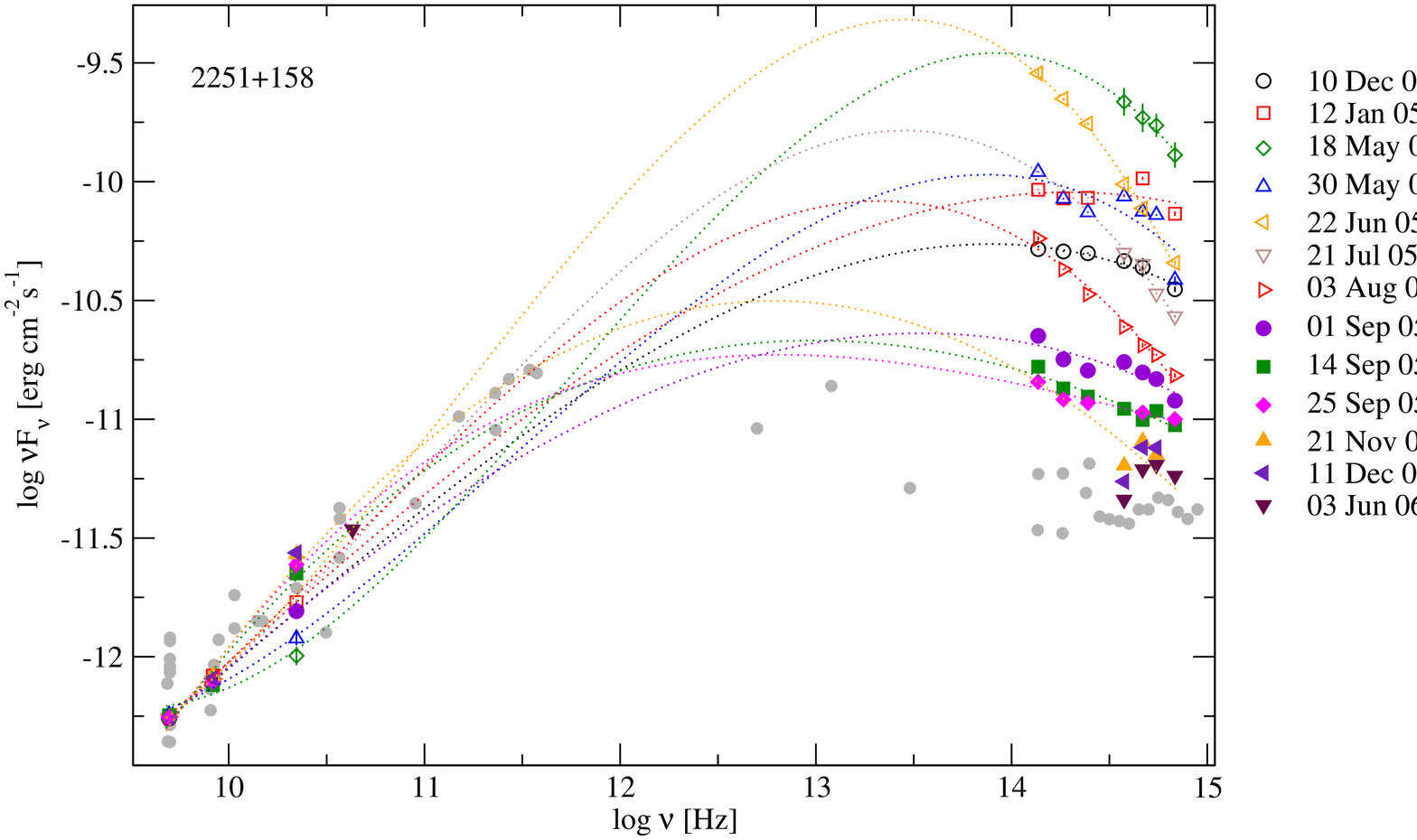}}
\caption{continued.}\label{fig:SEDs2}
\end{figure*}

\end{document}